\documentclass[aps,prc,reprint,superscriptaddress,floatfix]{revtex4-2}
\usepackage{graphicx}
\usepackage{float}
\usepackage{caption,subcaption}
\usepackage{dcolumn}
\usepackage{bm}
\usepackage{amsmath}
\usepackage{multirow}
\usepackage{xcolor}
\usepackage{hyperref}

\definecolor{mycrimson_html}{HTML}{DC143C}

\hypersetup{
    colorlinks=true,
    linkcolor=blue,
    citecolor=blue,
    urlcolor=mycrimson_html 
}

\begin{document}


\title{Microscopic study of the asymptotic behavior of the reduced width amplitude in $^7$Li and $^7$Be}

 \author{H. J. Zhu}
 \affiliation{College of Physics, Nanjing University of Aeronautics and Astronautics, Nanjing 210016, China}

 \author{M. J. Lyu}
 \affiliation{College of Physics, Nanjing University of Aeronautics and Astronautics, Nanjing 210016, China}

 \author{Q. Zhao}%
 \email{Corresponding author: zhaoqing91@zjhu.edu.cn}
 \affiliation{School of Science, Huzhou University, Huzhou 313000, Zhejiang, China}
 
 \author{Z. Cheng}
 \affiliation{College of Physics, Nanjing University of Aeronautics and Astronautics, Nanjing 210016, China}

  \author{J. Q. Tian}
 \affiliation{College of Physics, Nanjing University of Aeronautics and Astronautics, Nanjing 210016, China}

 \author{M. Kimura}
 \affiliation{Nuclear Reaction Data Centre (JCPRG), Hokkaido University, Sapporo 060-0810, Japan}
 \affiliation{Department of Physics, Hokkaido University, Sapporo 060-0810, Japan}
 \affiliation{RIKEN Nishina Center for Accelerator-based Science, RIKEN, Wako 351-0198, Japan}

 \author{T. Myo}
 \affiliation{General Education, Faculty of Engineering, Osaka Institute of Technology, 535-8585, Osaka, Japan}
 \affiliation{Research Center for Nuclear Physics (RCNP), Osaka University, 567-0047, Osaka, Japan}

 \author{H. Horiuchi}
 \affiliation{Research Center for Nuclear Physics (RCNP), Osaka University, 567-0047, Osaka, Japan}

 \author{H. Toki}
 \affiliation{Research Center for Nuclear Physics (RCNP), Osaka University, 567-0047, Osaka, Japan}

 \author{M. Isaka}
 \affiliation{Hosei University, 2-17-1 Fujimi, Chiyoda-ku, 102-8160, Tokyo, Japan}

 \author{H. Takemoto}
 \affiliation{Faculty of Pharmacy, Osaka Medical and Pharmaceutical University, 569-1094, Takatsuki, Osaka, Japan}
 
 \author{Akinobu Doté}
 \affiliation{KEK Theory Center, Institute of Particle and Nuclear Studies (IPNS), High Energy Accelerator Research Organization (KEK), 1-1 Oho, Tsukuba, Ibaraki, 305-0801, Japan}
 \affiliation{J-PARC Branch, KEK Theory Center, IPNS, KEK, 203-1, Shirakata, Tokai, Ibaraki, 319-1194, Japan}
 \affiliation{Graduate Institute for Advanced Studies, SOKENDAI, 1-1 Oho, Tsukuba, Ibaraki, 305-0801, Japan}

 \author{N. Wan}
 \affiliation{School of Physics and Optoeletronics, South China University of Technology, 510641, Guangzhou, China}

\begin{abstract}
We investigate the effects of different basis model spaces on the calculation of reduced width amplitude (RWA) and asymptotic normalization coefficient (ANC) for the $^{7}$Li and $^{7}$Be nuclei. The two-cluster model ($\alpha+t/^3$He) and three-cluster model ($\alpha+d+n/p$) with the generator coordinates method (GCM) are applied to calculate the wave function of $^7$Li and $^7$Be. Specifically, the model space for the three-cluster model is constructed upon a sufficiently broad space of two-cluster configuration bases by further including three-cluster configuration bases. We compare the impact on the results from two basis sets for these added three-cluster bases: one with a compact and one with a broad spatial distribution. The final results reveal that the two-cluster model cannot accurately reproduce the binding energies of $^7$Li and $^7$Be, and tends to overestimate their ANCs. Regarding the calculations with the three-cluster model, while the two basis sets do not give significant differences in energy or energy spectrum, the basis set with a compact model space fails to describe the asymptotic behavior of the RWA adequately. This introduces excessive uncertainty into the ANC calculation. In the end, we conclude that when calculating ANC via a microscopic framework, particular attention must be paid to ensuring sufficient model space, especially for the components describing the breakup channels. This approach provides ANC values for $^7$Li and $^7$Be that agree well with experimental results.

\end{abstract}

\maketitle

\section{Introduction}

The radiative capture reactions ${}^3\text{H}({}^4\text{He},\gamma){}^7\text{Li}$ and ${}^3\text{He}({}^4\text{He},\gamma){}^7\text{Be}$ are of significant importance in nuclear astrophysics. These two processes play a crucial role in determining the primordial abundance of ${}^7\text{Li}$ during Big Bang Nucleosynthesis (BBN)~\cite{RevModPhys.88.015004}. Furthermore, the ${}^3\text{He}({}^4\text{He},\gamma){}^7\text{Be}$ reaction is a key component of the solar proton-proton chains, initiating the branches responsible for the production of high-energy solar neutrinos~\cite{RevModPhys.70.1265}. These reactions are relevant to many frontier scientific challenges, such as the ``cosmological lithium problem'' and the ``solar neutrino puzzle''. The cosmological lithium problem refers to the persistent discrepancy where standard BBN theory predicts a primordial ${}^7\text{Li}$ abundance approximately three times higher than observed in ancient stars~\cite{PhysRevD.61.123505}. The solar neutrino puzzle, historically characterized by a conflict between predicted and measured solar neutrino fluxes, was ultimately resolved by the discovery of neutrino oscillations~\cite{RevModPhys.83.195}. Consequently, a precise determination of the reaction rates for these two processes in astrophysically relevant energy regions (approximately 20--500 keV for BBN and around 20 keV for the Sun) is critical for addressing these challenges. However, it is precisely in these important low-energy regions that the reaction cross sections become exceedingly small, making direct measurement exceptionally challenging and necessitating theoretical models~\cite{RevModPhys.88.015004,RevModPhys.70.1265,Pizzone_2014}.

At astrophysically relevant low energies, the dominant mechanism for radiative capture is often characterized as an external capture process. This is due to the existence of the Coulomb barrier, which restricts the reaction to occur primarily at radial distances beyond the range of nuclear interactions~\cite{PhysRevD.61.123505,RevModPhys.83.195,CHRISTY196189}. Under these conditions, the reaction cross section is predominantly determined by the asymptotic region of the nuclear wave function. The asymptotic normalization coefficient (ANC), which quantifies the amplitude of the bound state wave function in this region, is therefore a fundamental quantity for determining these capture reaction rates~\cite{PhysRevLett.73.2027,PhysRevC.67.065804}. Traditionally, the ANC is considered to be insensitive to the internal structure of the central part of the nucleus. Previous theoretical investigations treated these low-energy reactions as an approximate external capture process, employing simple two-body models that successfully reproduced the energy dependence of the measured cross sections~\cite{PhysRevC.48.1420,PhysRevC.79.065804,1995PAN....58..579D,CHRISTY196189,PhysRevC.23.645,LANGANKE1986351,KAJINO1986559}.

However, recent studies indicate that the ANC's sensitivity extends beyond traditional assumptions. Studies using the microscopic cluster model~\cite{MERTELMEIER1986387,csoto2000study} find that moving beyond the single-channel approximation by incorporating the (${}^6\text{Li} + n/p$) channel, the extension of the model space alters not only the absolute value of the S-factor, but also significantly modifies its low-energy slope. These studies suggest that a simple two-cluster model may overlook the effects of more complex internal nuclear structures on the precise prediction of reaction cross sections and nuclear structural properties. Advanced studies~\cite{NKTimofeyuk_2008,Timofeyuk_2014} also suggest that valence nucleon correlations in three-body cluster systems (e.g., ${}^{12}\text{Be}\rightarrow{}^{10}\text{Be}+n+n$) induce non-standard asymptotic behavior in the wave function. These findings suggest that the conventional assumption of a pure external capture may not be entirely valid, even at low energies, and that the complex internal structure of the nucleus can exert a persistent influence on the asymptotic behavior of the wave function.

Motivated by these findings, we aim to explore the impact of using a many-body model on the estimation of ANC for  $^7$Li(g.s.)$\rightarrow\!\alpha +t$ and $^7$Be(g.s.)$\rightarrow\!\alpha +{}^3\text{He}$. Within a microscopic many-body framework, we systematically investigate the differences in their asymptotic behavior when considering a purely two-cluster model versus those including three-cluster configurations. We also investigate the effect of using basis sets with different spatial distributions on the diagonalization of wave functions for ANC results. By comparing these results from different model approaches, we seek to provide a more accurate and comprehensive understanding of the role of nuclear structure in determining ANC.

This article is organized as follows. The theoretical frameworks, including the microscopic method combining a neural network for calculating the wave function, are detailed in Sec.~II. Results and discussion concerning the energy spectra and ANCs are presented in Sec.~III. Finally, the conclusions are summarized in Sec.~IV.

\section{THEORETICAL FRAMEWORK}

\subsection{The Hamiltonian and the wave function}
In Ref.~\cite{10.1093/ptep/ptae187}, Myo et al. used the Multi-Cool method to calculate the wave functions of the Li isotopes and reproduce the energy spectra and radii well. The Hamiltonian and single-particle wave function parameters are adopted from Ref.~\cite{10.1093/ptep/ptae187}, as a similar microscopic method is employed in this work. The Hamiltonian adopted is as follows:
\begin{equation} \label{eq:hamiltonian}
\begin{split}
    \hat{\mathcal{H}} ={}& \sum_i^A \hat{t}_i - \hat{T}_{\text{c.m.}} + \sum_{i<j}^A \hat{v}_{ij}^{NN} \\
                         & + \sum_{i<j}^A \hat{v}_{ij}^{LS} + \sum_{i<j \in \text{protons}}^A \hat{v}_{ij}^{C},
\end{split}
\end{equation}
where $A$ denotes the nucleon number, $\hat{t}_i$ corresponds to the kinetic energy term, and $\hat{T}_{c.m.}$ represents the center-of-mass kinetic energy term. $\hat{v}_{ij}^{NN}$, $\hat{v}_{ij}^{LS}$, $\hat{v}_{ij}^{C}$ represent the $NN$ interaction, spin-orbit interaction, and Coulomb interaction terms, respectively. The $NN$ interaction is adopted as the Volkov No.2 potential~\cite{VOLKOV196533}, whose functional form is given by:
\begin{equation}
\begin{split}
    \hat{v}_{ij}^{NN} ={}& \left(W - M\hat{P}^\sigma\hat{P}^\tau+B\hat{P}^\sigma-H\hat{P}^\tau\right) \\
                         & \times \left[V_1\exp\left(-r_{ij}^2/c_1^2\right) + V_2\exp\left(-r_{ij}^2/c_2^2\right)\right].
\end{split}
\end{equation}
The spin-orbit interaction term $\hat{v}_{ij}^{LS}$ adopts the G3RS potential~\cite{10.1143/PTP.39.91,10.1143/PTP.62.1018}:
\begin{equation}
\hat{v}_{ij}^{LS}=V_{ls}\left(e^{-d_{1}\mathbf{r}_{ij}^{2}}-e^{-d_{2}\mathbf{r}_{ij}^{2}}\right)\hat{P}_{31}\hat{L}\cdot\hat{S}
\label{eq:4}
\end{equation}
Here $\hat{P}_{31}$ projects the two-cluster system into a triplet odd state.

The wave function is constructed using the Generator Coordinate Method (GCM) within a cluster model framework. In this method, the basis wave function is constructed by the cluster wave function with antisymmetrization as
\begin{equation}
\Phi=\mathcal{A}\{\psi_{c_1}(\mathbf{Z}_{c_1})\psi_{c_2}(\mathbf{Z}_{c_2})...\}~.
\end{equation}
Here $c_i$ denotes the $i$th cluster in the nucleus, and all the nucleons in it share the same generator coordinate $\mathbf{Z}_{c_i}$. The cluster wave function is also a Slater determinant defined by the single-particle wave functions as
\begin{equation}
\psi_c=\mathcal{A}\{\phi_{1}(\mathbf{Z}_{c})\phi_{2}(\mathbf{Z}_{c})...\}~,
\end{equation}
where the single-particle wave function is defined in the Gaussian form with spin and isospin parts:
\begin{equation}
\phi(\mathbf{r},\mathbf{Z})=\left(\frac{2\nu}{\pi}\right)^{3/4}\exp\left[-\nu\left(\mathbf{r}-\frac{\mathbf{Z}}{\sqrt{\nu}}\right)^2+\frac{1}{2}\mathbf{Z}^2\right]\chi\tau~.
\end{equation}
The spin and isospin parts are $\chi = a\mid\uparrow\rangle + b\mid\downarrow\rangle$, and $\tau = \text{proton or neutron}$. The generator coordinate $\mathbf{Z}$ is a complex vector defined as $\mathbf{Z}=\mathbf{R}+i\mathbf{D}$. The real part represents the spatial coordinate and the imaginary part represents the momentum of the particle. Each basis wave function is defined by a set of generator coordinates. In the single-cluster model, the generator coordinates of all nucleons are fixed at the origin. The two-cluster model divides the nucleus into two parts, arranging them along a straight line with equal spacing to construct the system's basis vectors. For a three-cluster system, a newly proposed control neural network (Ctrl.NN) is used to optimize the position of each cluster. This network uses the system’s ground state energy as a loss function to find the optimal cluster arrangement by minimizing it~\cite{CHENG2025139397,TIAN2024138816}. This method utilizes a neural network to optimize multiple basis wave functions simultaneously for the intrinsic energy and can construct a larger model space. 

The total wave function with definite angular momentum and parity is expressed as a superposition of projected basis states:
\begin{equation}
\Psi = \sum_i g_{i,K}\hat{P}^{J^\pi}_{MK}\Phi_i~.
\end{equation}
where $\hat{P}^{J^\pi}_{MK}$ is the projection operator. The coefficients $g_{i,K}$ and the eigenenergies are obtained by solving the Hill-Wheeler equation.

\begin{figure}
  \centering
    \includegraphics[width=0.48\textwidth]{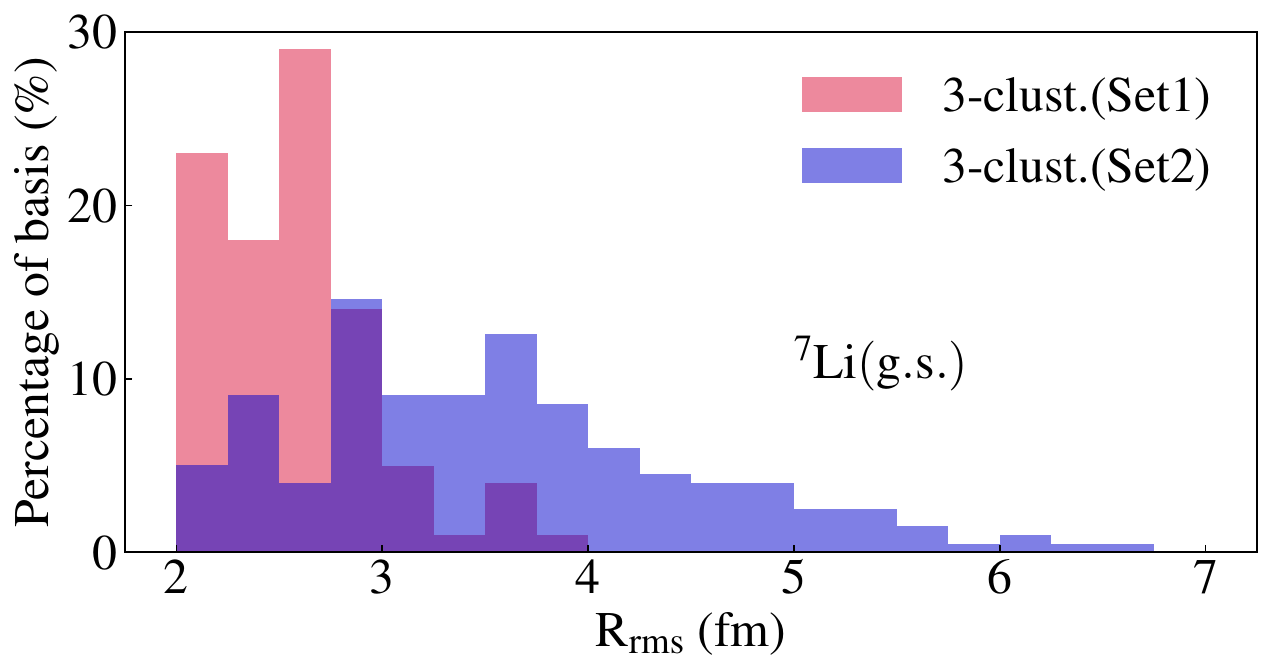}
    \includegraphics[width=0.48\textwidth]{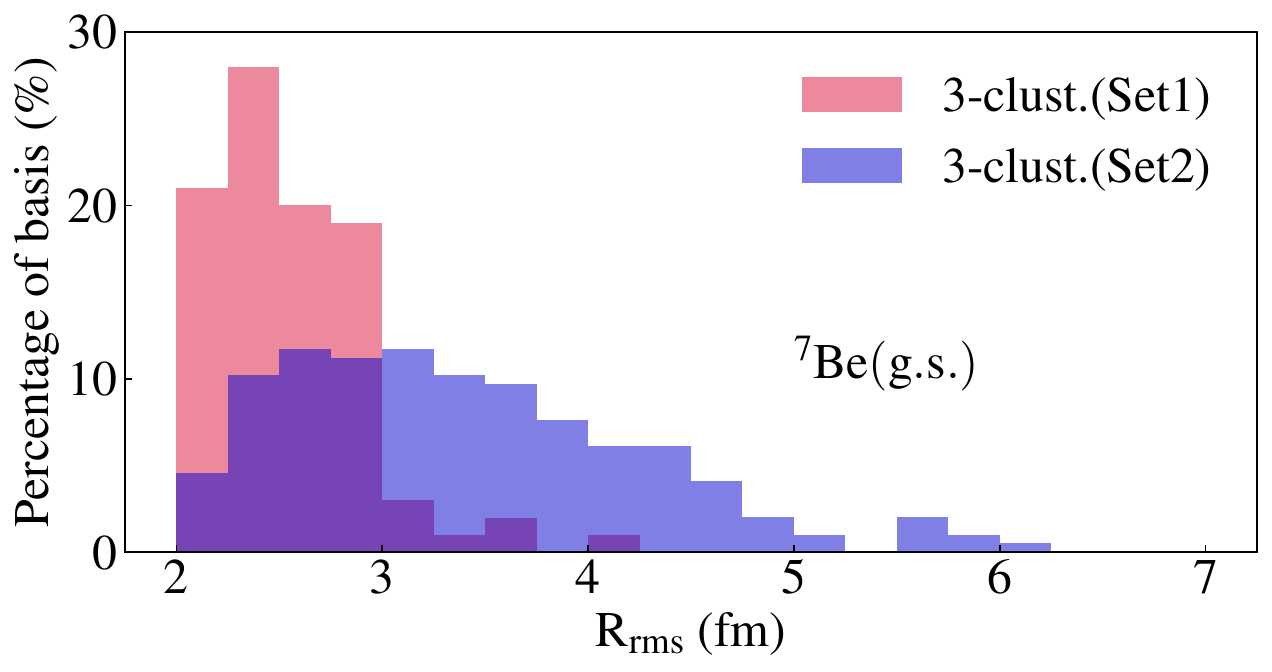}
    \caption{\label{fig:rms} Distribution of the r.m.s. radii for the three-cluster configuration component of the basis set. The vertical axis represents the percentage of the total number of basis wave functions. }
\end{figure}

The spatial structures of $^{7}$Li and $^7$Be can be regarded as two types of cluster models. One is $\alpha + t/{}^3\text{He}$, and the other is $\alpha+d+n/p$. The corresponding basis wave functions are constructed as
\begin{equation}
\Phi=\mathcal{A}\{\psi_{\alpha}(\mathbf{Z}_{\alpha})\psi_{t/^3He}(\mathbf{Z}_{t/^3He})\}
\end{equation}
and
\begin{equation}
\Phi=\mathcal{A}\{\psi_{\alpha}(\mathbf{Z}_{\alpha})\psi_{d}(\mathbf{Z}_{d})\psi_{n/p}(\mathbf{Z}_{n/p})\}~.
\end{equation}
These are referred to as the two-cluster (2-clust.) and three-cluster (3-clust.) models, respectively. The three-cluster (3-clust.) model space is constructed by augmenting the full basis set of the two-cluster (2-clust.) model with an additional set of three-cluster configuration basis states. Therefore, the model space of the two-cluster model is fully contained within that of the three-cluster model. We also test two sets for the three-cluster model. Set1 consists of three-cluster configuration basis states with a spatially localized distribution for the generated coordinates, whereas Set2 features a more widespread distribution. Both sets are generated by initializing the basis states in different spatial regions, followed by an optimization process guided by a control neural network. In Fig~\ref{fig:rms}, we show the distribution of the root-mean-square (r.m.s.) radius of the basis wave function for both Set1 and Set2.
It can be seen that the basis in Set1 mostly has compact r.m.s. radii, while those in Set2 exhibit a broader distribution of r.m.s. radii. In this work, we aim to test the impact of employing these different structural models on the calculated ANC results when constructing the wave function.

\subsection{Reduced Width Amplitude and Asymptotic Normalization Constant}

The reduced width amplitude (RWA) represents the wave function of a subsystem in a nucleus, which is defined as the overlap between the wave function of the total nucleus and the subsystem constructed by two residues.
\begin{equation}
a y_{l}(a) = a \sqrt{\frac{A!}{(1+\delta_{g_{1}g_{2}})g_{1}!g_{2}!}} \left\langle \frac{\delta(r-a)}{r^{2}} \Psi_{c_{1}}\Psi_{c_{2}}Y_{l}(\hat{r}) \middle| \Psi_{A} \right\rangle~.
\end{equation}
Here $\Psi_{c_1}$ and $\Psi_{c_2}$ are the wave functions of the clusters with the number of nucleons $g_1$ and $g_2$. $\Psi_A$ is the wave function of the total nucleus. To ensure consistency across our models, the structures used to calculate the cluster wave functions correspond to the structure used for the parent nucleus. The term $l$ represents the relative orbital angular momentum between these clusters. In this work, the RWA is computed via the Laplace expansion method~\cite{10.1093/ptep/ptx063}.

At large separations between the two residues, where $NN$ interaction becomes negligible and only the Coulomb interaction persists, the behavior of the RWA takes specific forms depending on whether the state is bound.

For bound states ($E < 0$), the solutions of the Schr\"odinger equation at large separations should follow the Whittaker function $W_{\kappa,\mu}(z)$ with a normalization constant $C$.
\begin{equation}
a y_l(a) = C W_{\kappa,\mu}(2ka)~.
\label{eq:16}
\end{equation}
$k=\sqrt{-2m E/\hbar^2}$, where $m$ is the reduced mass. The parameters are $\kappa = -Z_1 Z_2 e^2 m / (\hbar^2 k)$ and $\mu = l + 1/2$. The factor $C$ is called the asymptotic normalization constant (ANC), a quantity characterizing the virtual decay of a nucleus into two parts, which is of importance for the study of astrophysical reactions. In the asymptotic region where $C$ becomes constant, the logarithmic derivative of the calculated RWA must match that of the Whittaker~\cite{PhysRevC.106.054313}.
\begin{equation}
\label{eq:dlog}
\frac{[a y_l(a)]'}{a y_l(a)} = \frac{W'_{\kappa,\mu}(2ka)}{W_{\kappa,\mu}(2ka)}~.
\end{equation}
By comparing the RWA with the Whittaker function, we can analyze its asymptotic behavior and determine the ANC.

\section{RESULTS AND DISCUSSIONS}

\subsection{Results and model space configuration}

Based on the variational principle, the calculated energy serves as the fundamental metric for verifying the wave function. We firstly show the ground state energies of the corresponding nuclei in Table~\ref{tab:ene} and compare them with the experimental data. The charge radii of $^7$Li and $^7$Be are also listed.
\begin{table}
\caption{\label{tab:ene}
Experimental values and calculated results for the ground state energies ($E$) and charge radii ($R_c$) of various nuclei. The abbreviation ``clust." stands for cluster. Experimental data are from Refs.~\cite{PURCELL20151,TILLEY19921,TILLEY20023}.
}
\renewcommand{\arraystretch}{1.2}
\begin{ruledtabular}
\begin{tabular*}{\columnwidth}{@{\extracolsep{\fill}}lcccc}
\textrm{E} & \textrm{Expt.} & \textrm{2-clust.} & \textrm{3-clust.(set1)} & \textrm{3-clust.(set2)} \\
\colrule
$E(\mathrm{^3H})$ (MeV) & -8.48 & -6.88 & -8.53 & -8.53 \\
$E(\mathrm{^3He})$ (MeV) & -7.72 & -6.12 & -7.88 & -7.88 \\
$E(\mathrm{^4He})$ (MeV) & -28.30 & -27.62 & -27.62 & -27.62 \\
$E(\mathrm{^7Li})$ (MeV) & -39.25 & -37.13 & -38.68 & -38.77 \\
$E(\mathrm{^7Be})$ (MeV) & -37.60 & -35.48 & -37.40 & -37.31 \\
\colrule
$R_c$ \\
\colrule
$R_c(\mathrm{^7Li})$ (fm) & 2.44 & 2.55 & 2.53 & 2.55 \\
$R_c(\mathrm{^7Be})$ (fm) & 2.65 & 2.75 & 2.73 & 2.76 \\
\end{tabular*}
\end{ruledtabular}
\end{table}
The main difference between 2-clust. and 3-clust. models is the breaking of $^3$H and $^3$He. Therefore, we also construct the wave functions of these two subsystems under the same concept for 2-clust. and 3-clust., respectively. In this table, it is not surprising that the energies calculated by the 3-clust. model are generally lower than those of the 2-clust. model, indicating that the 3-clust. model gives better wave functions. This is because of the clear advantages of the 3-clust. model in the model space. On the other hand, both for the 3-clust. model, Set1 and Set2 exhibit no significant difference in ground-state energy. Set2 possesses a slightly greater charge radius due to its larger spatial distribution.

\begin{figure}
  \centering
    \includegraphics[width=0.48\textwidth]{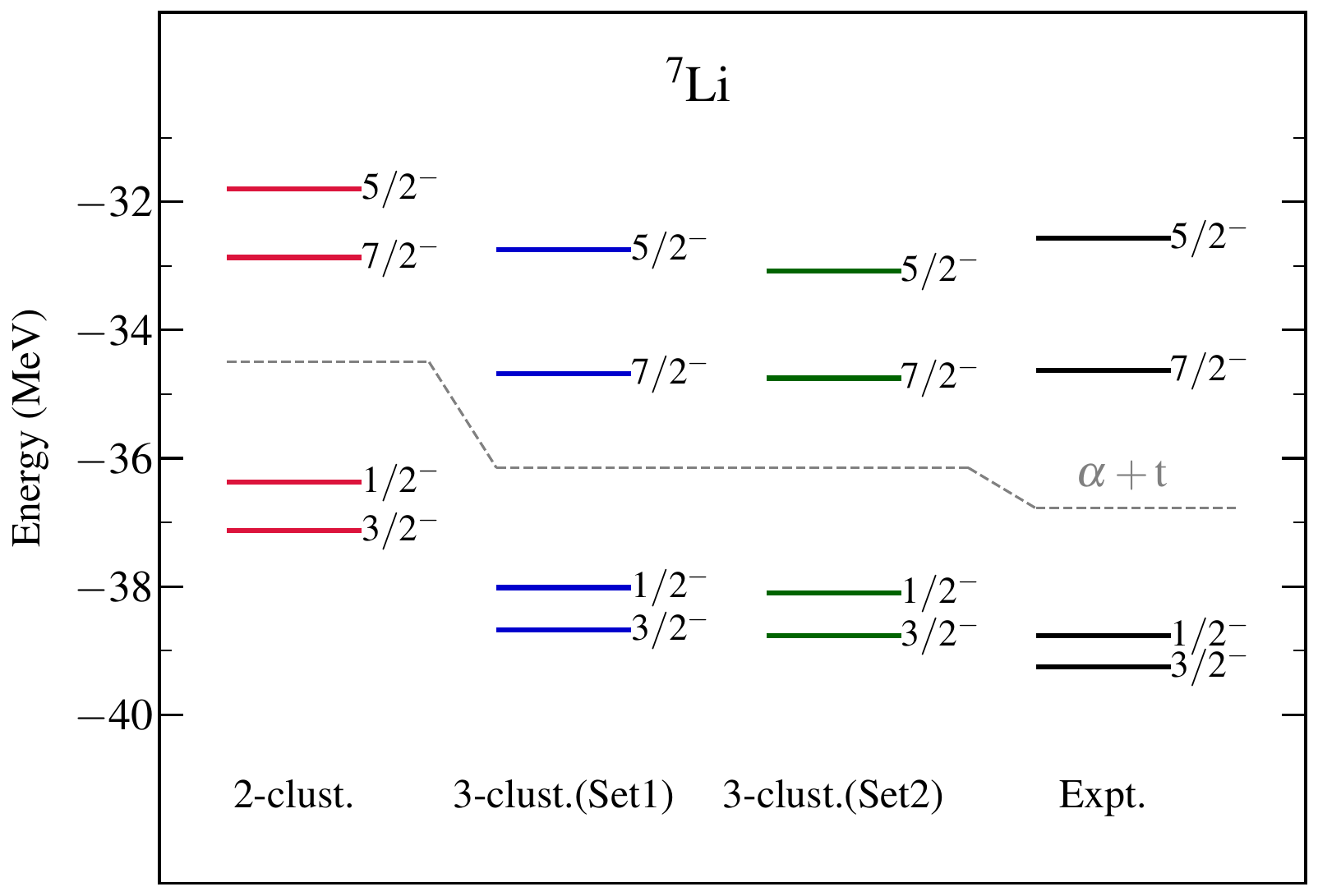}
    \includegraphics[width=0.48\textwidth]{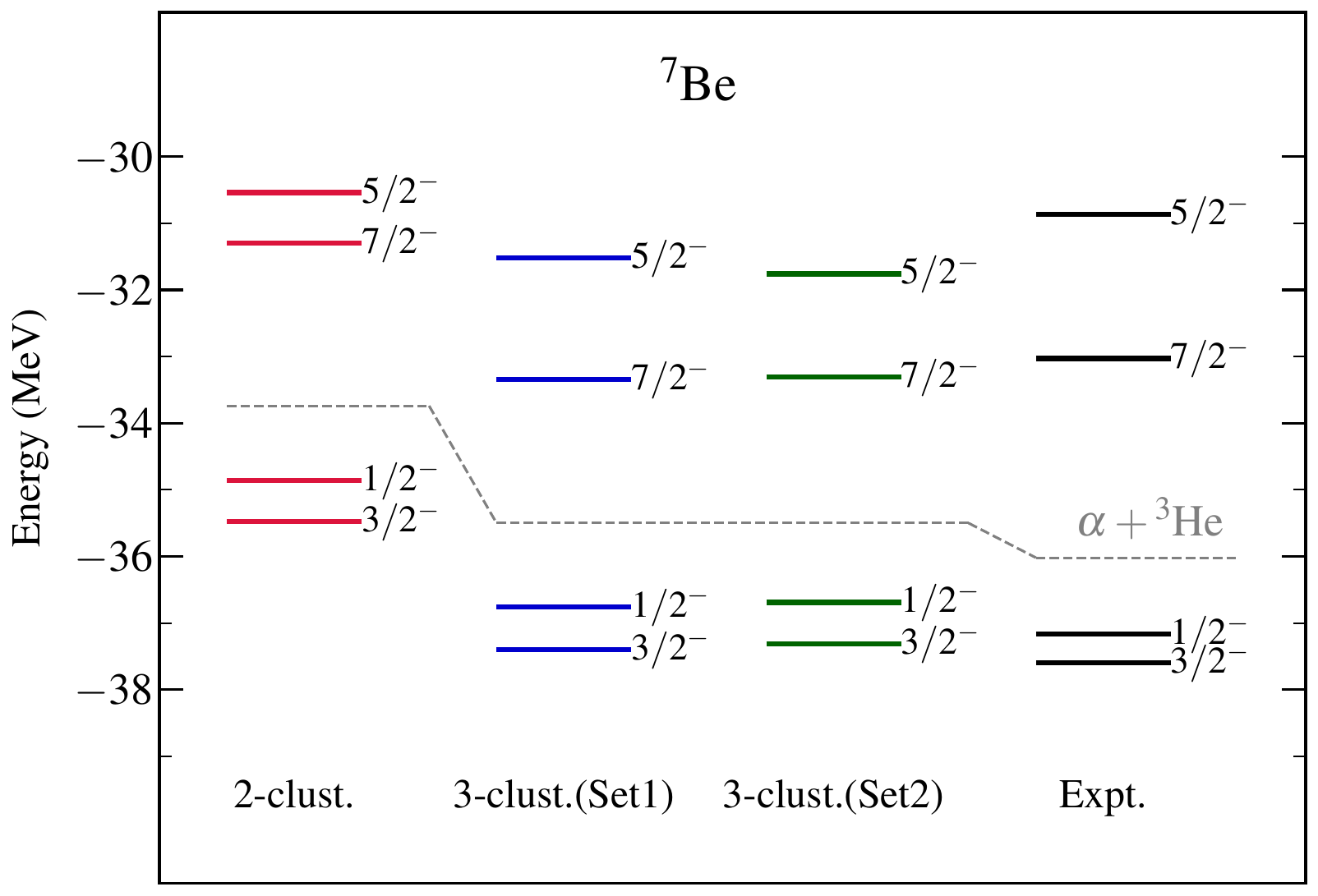}
    \caption{\label{fig:enespec} Energy spectra of $^{7}$Li and $^{7}$Be under different sets in comparison with the experiments~\cite{TILLEY20023}.}
\end{figure}

To further test the results of 2-clust., 3-clust.(Set1), and 3-clust.(Set2) in the eigen-states, we also present the energy spectrum results under three different sets, as shown in the Fig.~\ref{fig:enespec}.
Here, we again observe that the energies obtained from the 3-clust. model are generally lower than those from the 2-clust. model and closer to experimental values. Nevertheless, the 2-clust. model still gives excitation energy results that agree reasonably with experiments. For Set1 and Set2, we still observe no significant differences between them in the energy spectrum. In fact, we calculate the overlap between the ground-state wave functions of these two sets.
$$
O = \frac{|\langle\Psi_{\text{Set1}} | \Psi_{\text{Set2}}\rangle|^2}{\langle\Psi_{\text{Set1}} | \Psi_{\text{Set1}}\rangle \langle\Psi_{\text{Set2}} | \Psi_{\text{Set2}}\rangle}
$$
The overlap is $99.7\%$. This result implies that, in the usual sense, these two wave functions are nearly identical. Referring to the distribution of the r.m.s. radii of the basis in Fig~\ref{fig:rms}, we can conclude that under the Ctrl.NN method, both the calculations of Set1 and Set2 have converged. The basis in Set2 with larger r.m.s. radii will not further contribute to the final energy results. However, this conclusion likely applies only to microscopic calculations that typically focus on energy or internal structure of the nucleus. We will later see how these bases, which are easily overlooked in energy terms, significantly influence the calculations of ANC.

\subsection{RWA and its asymptotic behavior}

We then discuss the impact of these models on RWA results and their asymptotic behavior. As introduced in the framework section, the RWA is a wave function of a substructure in the total nucleus system. To obtain the ANC of $^7$Li and $^7$Be, we need the two-body RWA between the $\alpha$-particle and the residue nucleus, namely $^3$H ($t$) and $^3$He. It should be noted that although we apply 2-clust. or 3-clust. models in the wave function calculations, these models are formulated for the generated coordinate space. The obtained final wave function remains a microscopic many-body model wave function, which can then be used to compute the two-body RWA. 

\begin{figure}[htbp]
  \centering
  \includegraphics[width=0.48\textwidth]{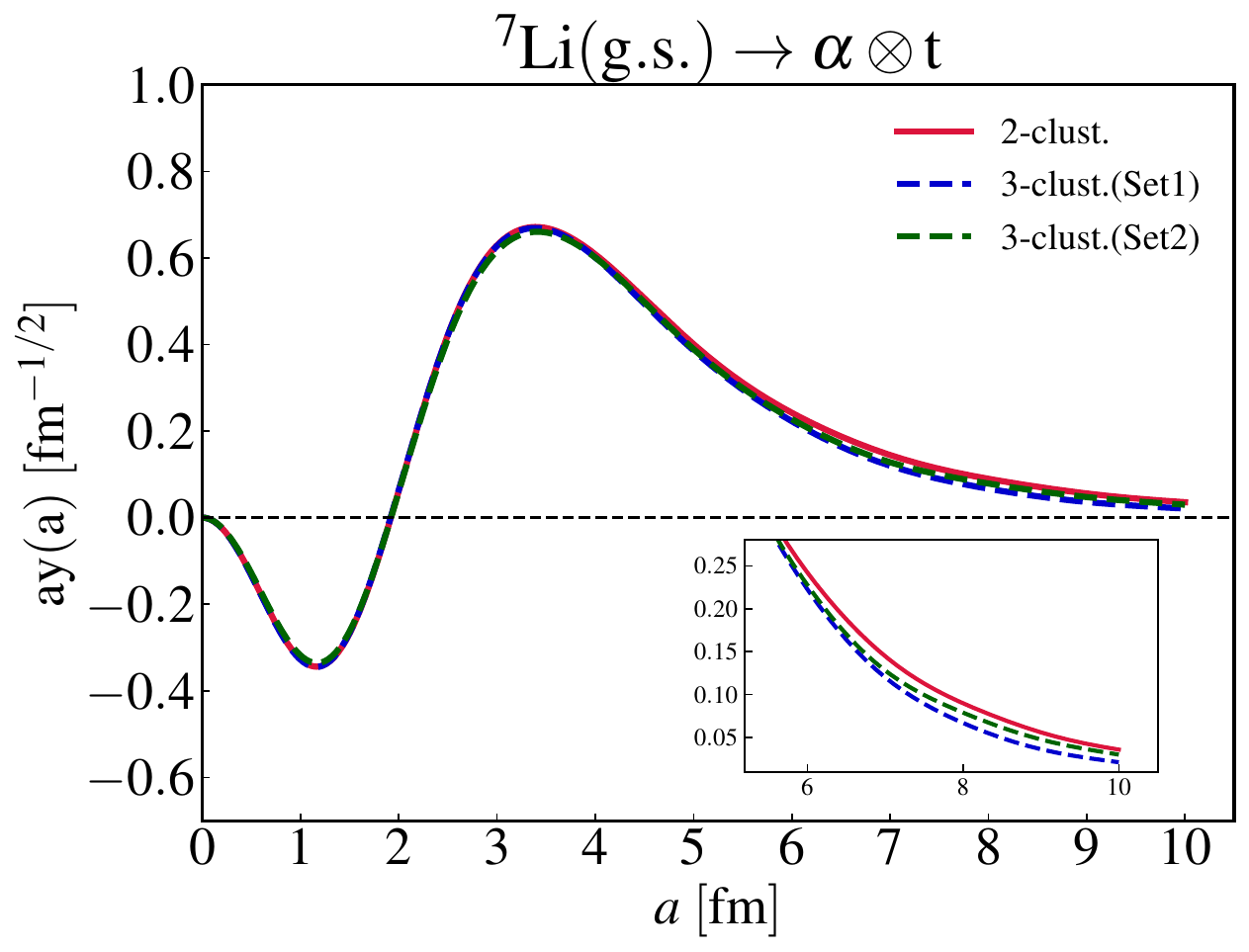}
  \includegraphics[width=0.48\textwidth]{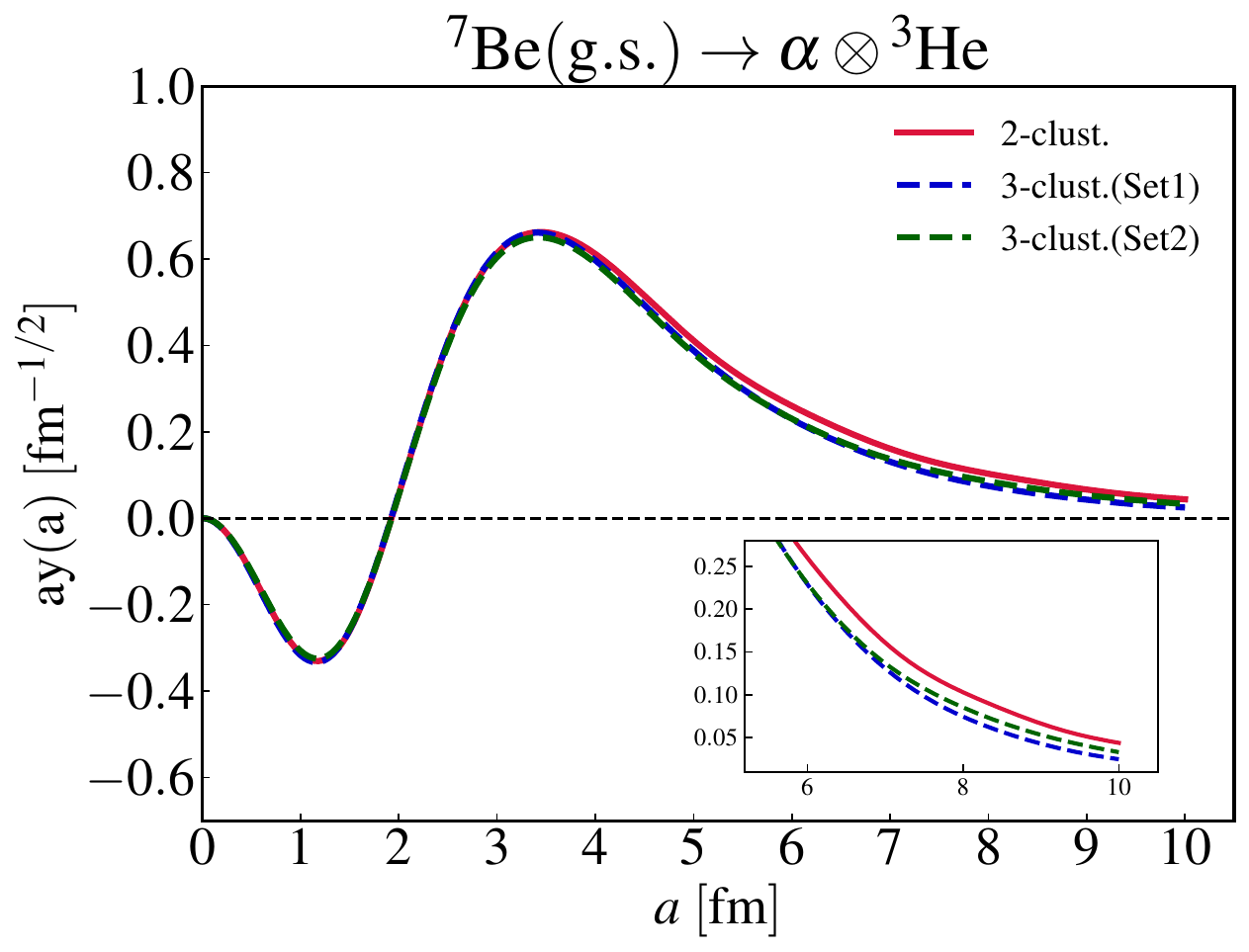}
  \caption{\label{fig:rwa} The RWAs calculated for $^7$Li(g.s.)$\rightarrow \alpha +t$ and $^7$Be(g.s.)$\rightarrow \alpha +{}^3\text{He}$ channels under different sets.}
\end{figure}

In Fig.~\ref{fig:rwa}, we compare the RWAs of 2-clust. and 3-clust. (Set1) with those of 3-clust. (Set2), 
where we find that their RWA results show no significant differences except for minor variations at the tails. It indicates that the internal structures of the nuclei described by these wave functions are largely similar. The tail regions over $6$ fm exhibit subtle differences, which means that their asymptotic behavior is not identical. 
\begin{figure}
  \centering
    \includegraphics[width=0.48\textwidth]{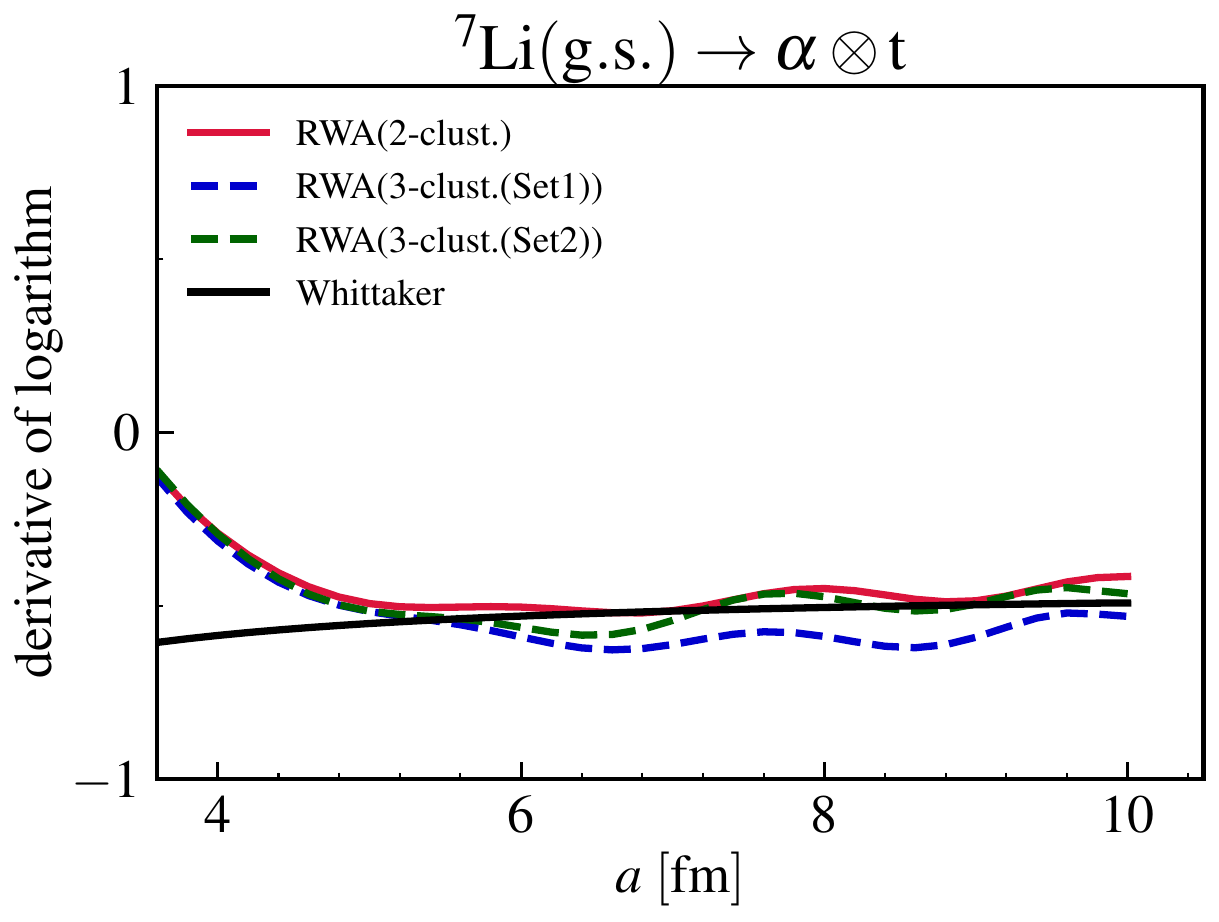}
    \includegraphics[width=0.48\textwidth]{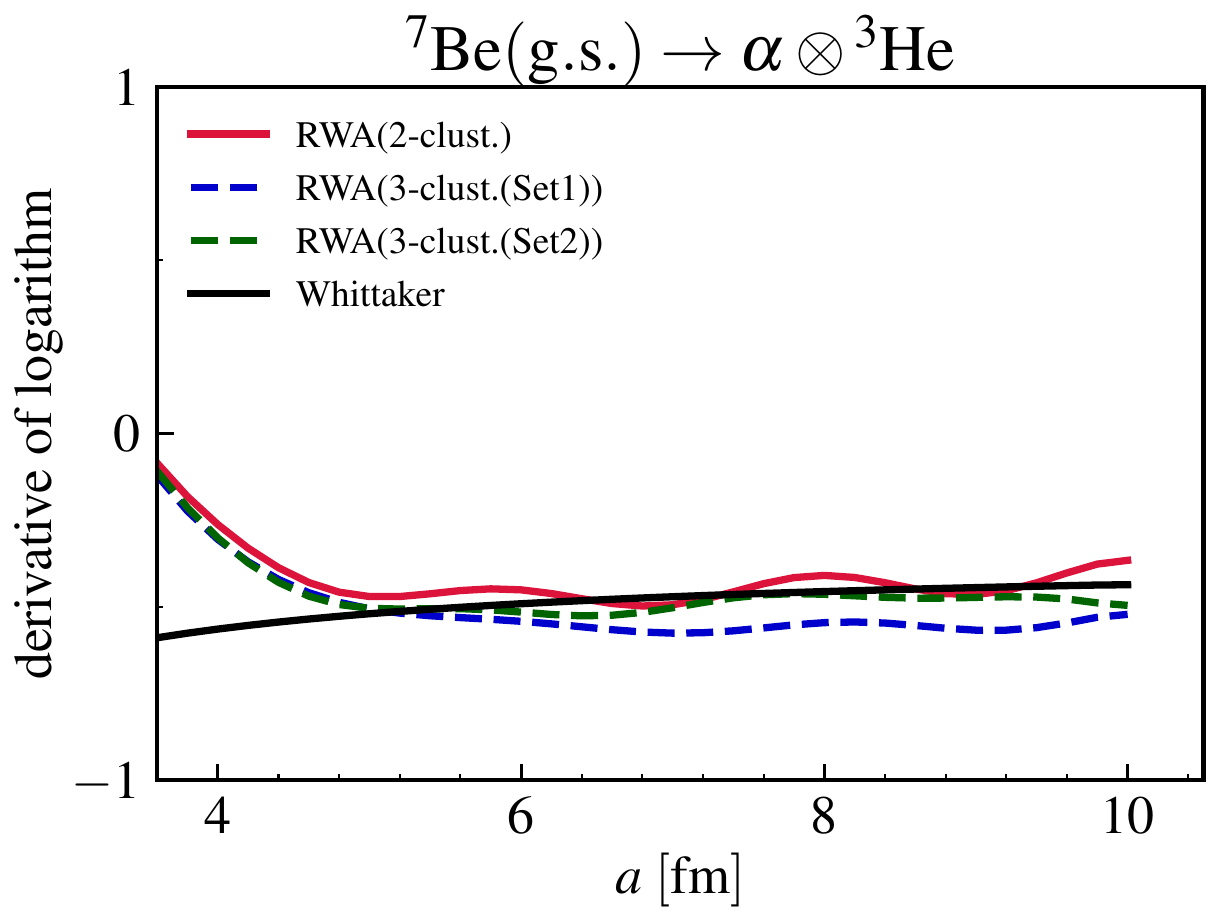}
    \caption{\label{fig:dlog} Comparison of the logarithmic derivatives of the RWA and Whittaker function under different sets.}
\end{figure}
\begin{figure*}[htbp]
  \centering
  \begin{subfigure}[b]{0.48\textwidth}
    \includegraphics[width=\linewidth]{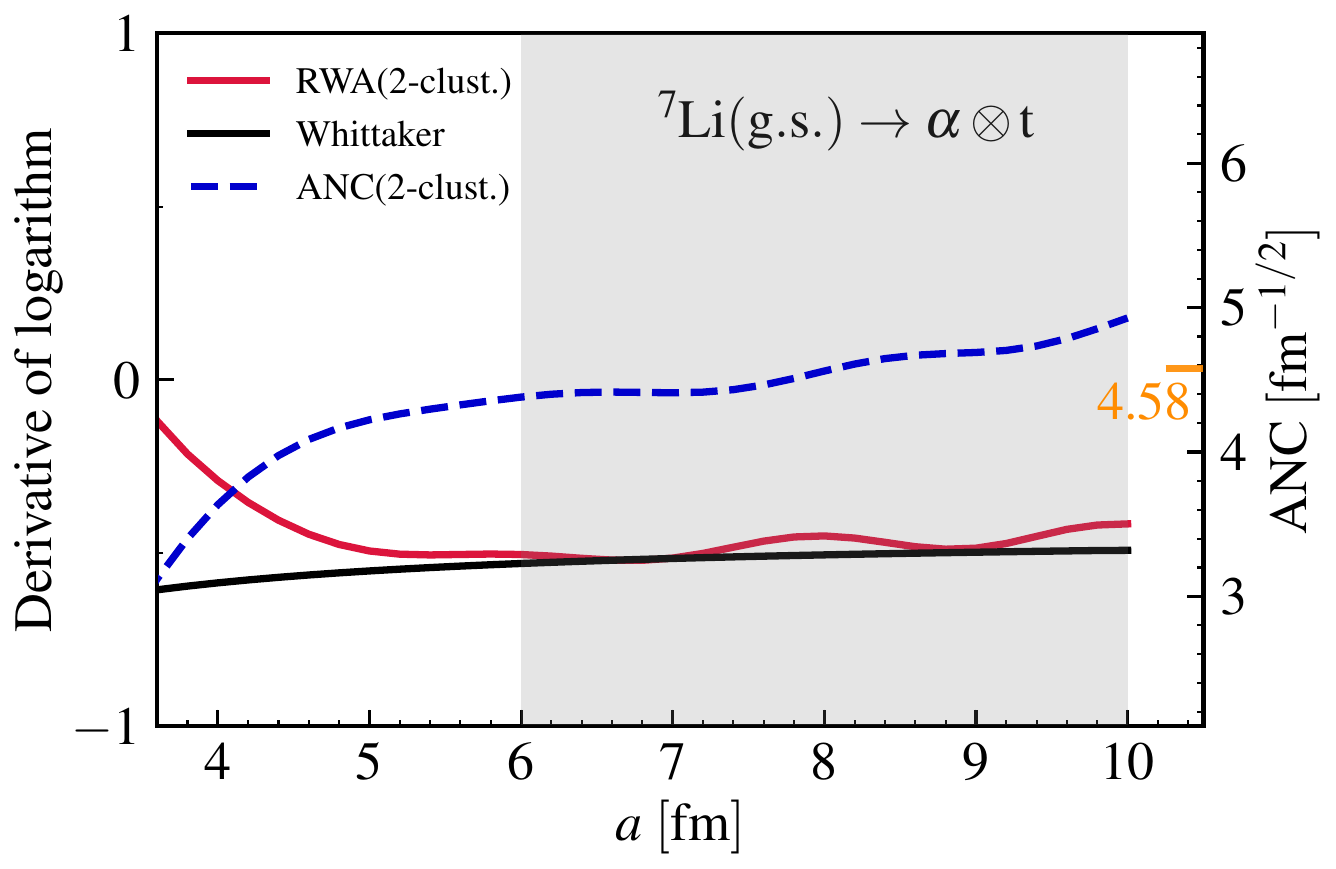}
  \end{subfigure}
  \hfill
  \begin{subfigure}[b]{0.48\textwidth}
    \includegraphics[width=\linewidth]{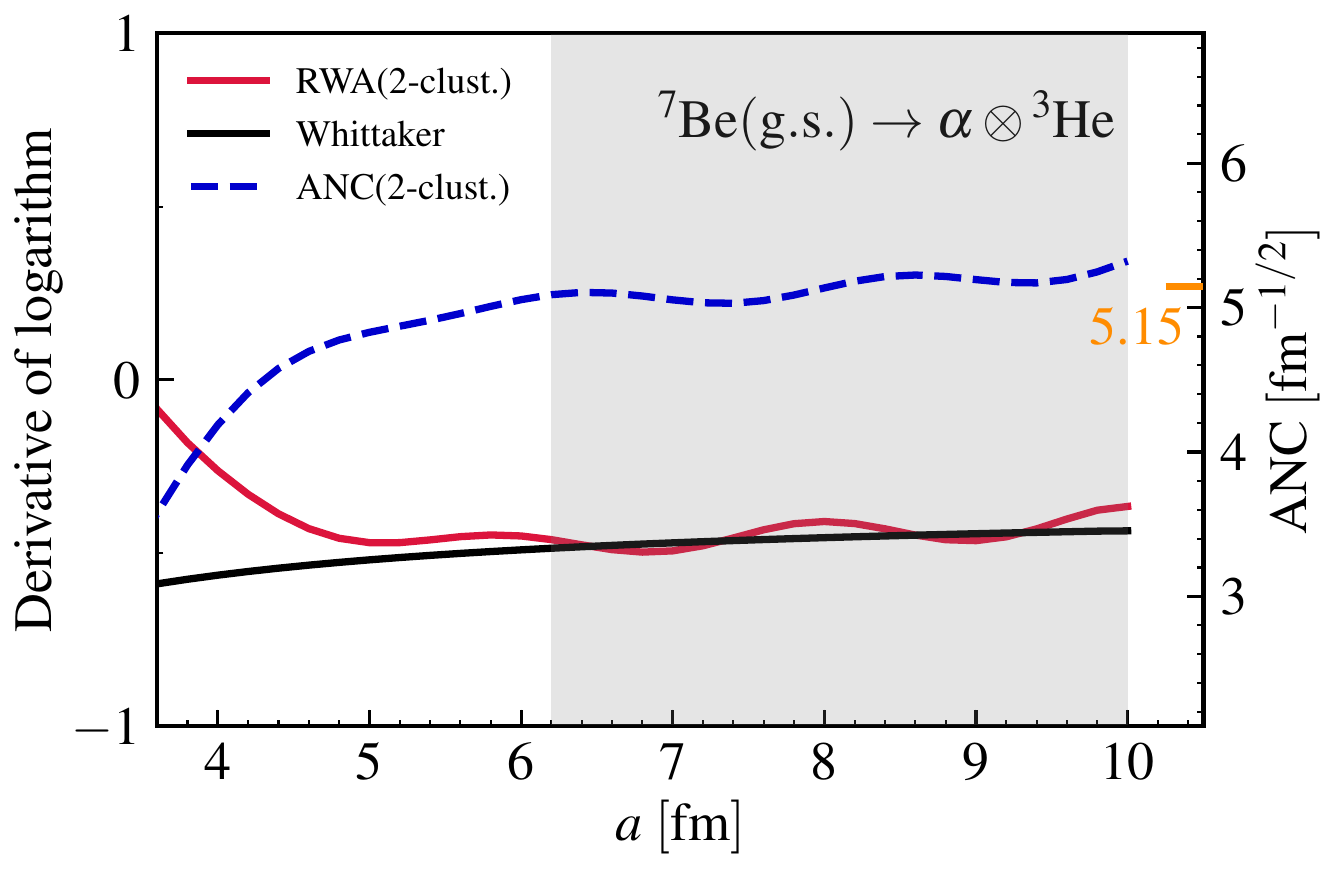}
  \end{subfigure}

  \vskip\baselineskip
  
  \begin{subfigure}[b]{0.48\textwidth}
    \includegraphics[width=\linewidth]{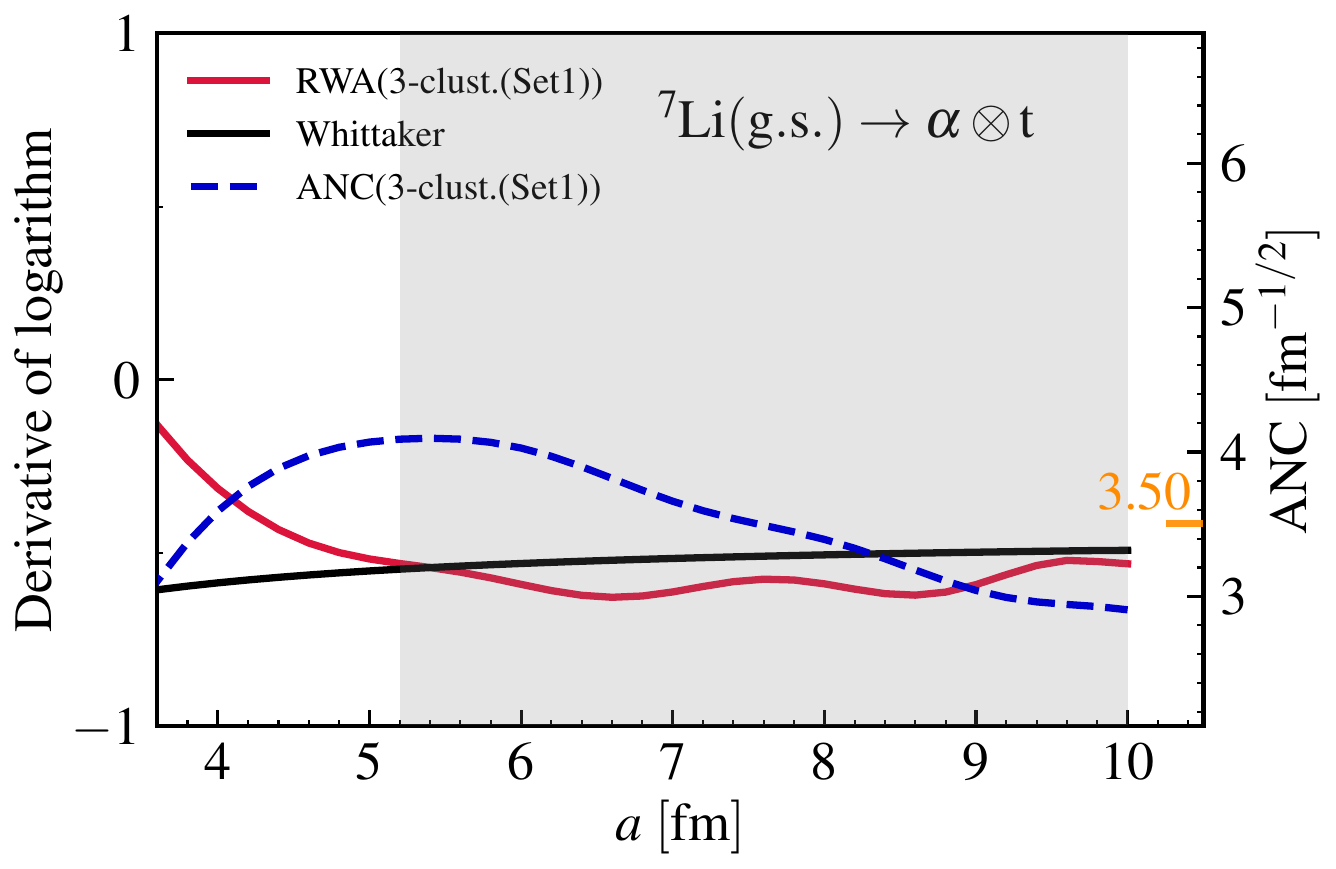}
  \end{subfigure}
  \hfill
  \begin{subfigure}[b]{0.48\textwidth}
    \includegraphics[width=\linewidth]{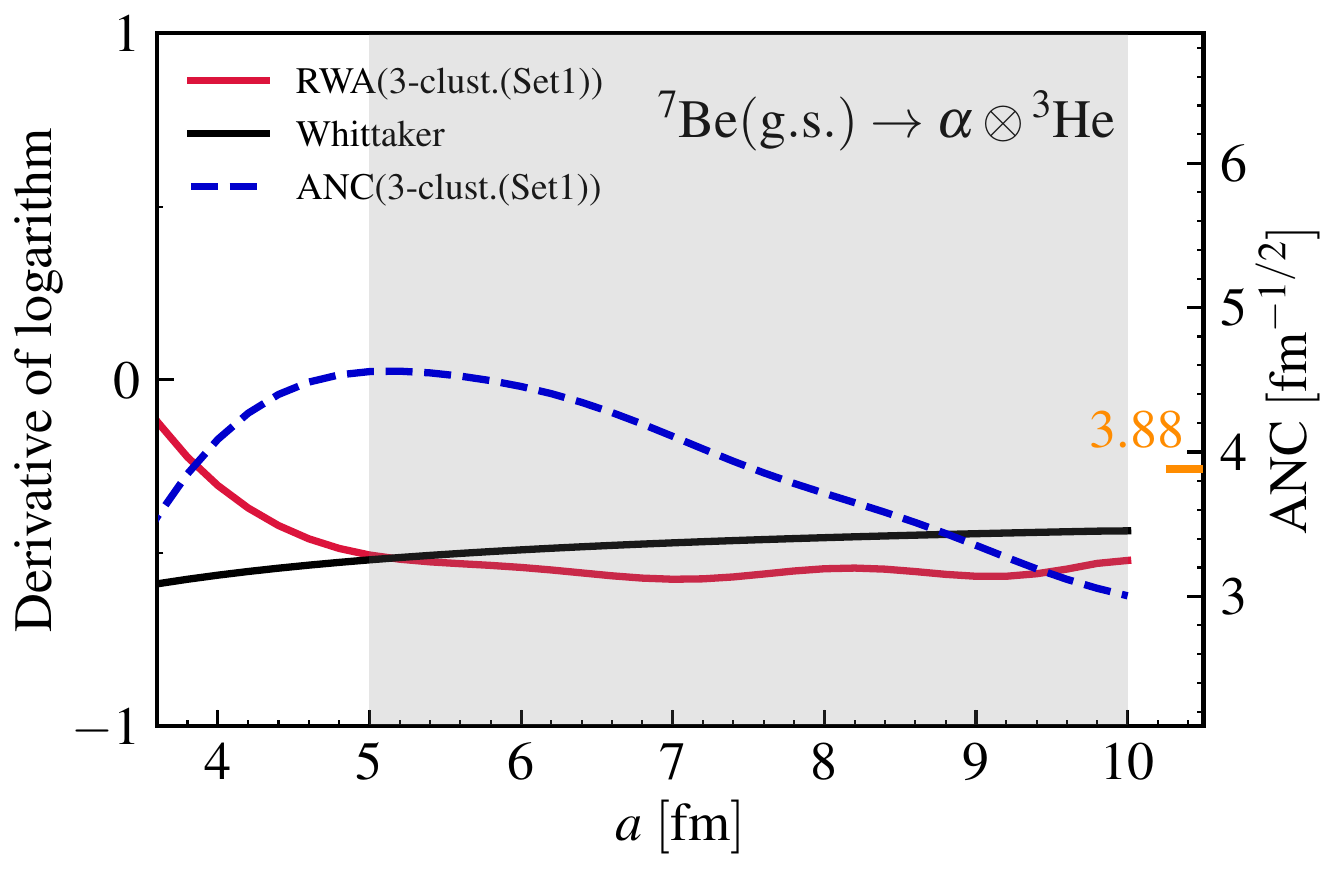}
  \end{subfigure}
  
  \vskip\baselineskip
  
  \begin{subfigure}[b]{0.48\textwidth}
    \includegraphics[width=\linewidth]{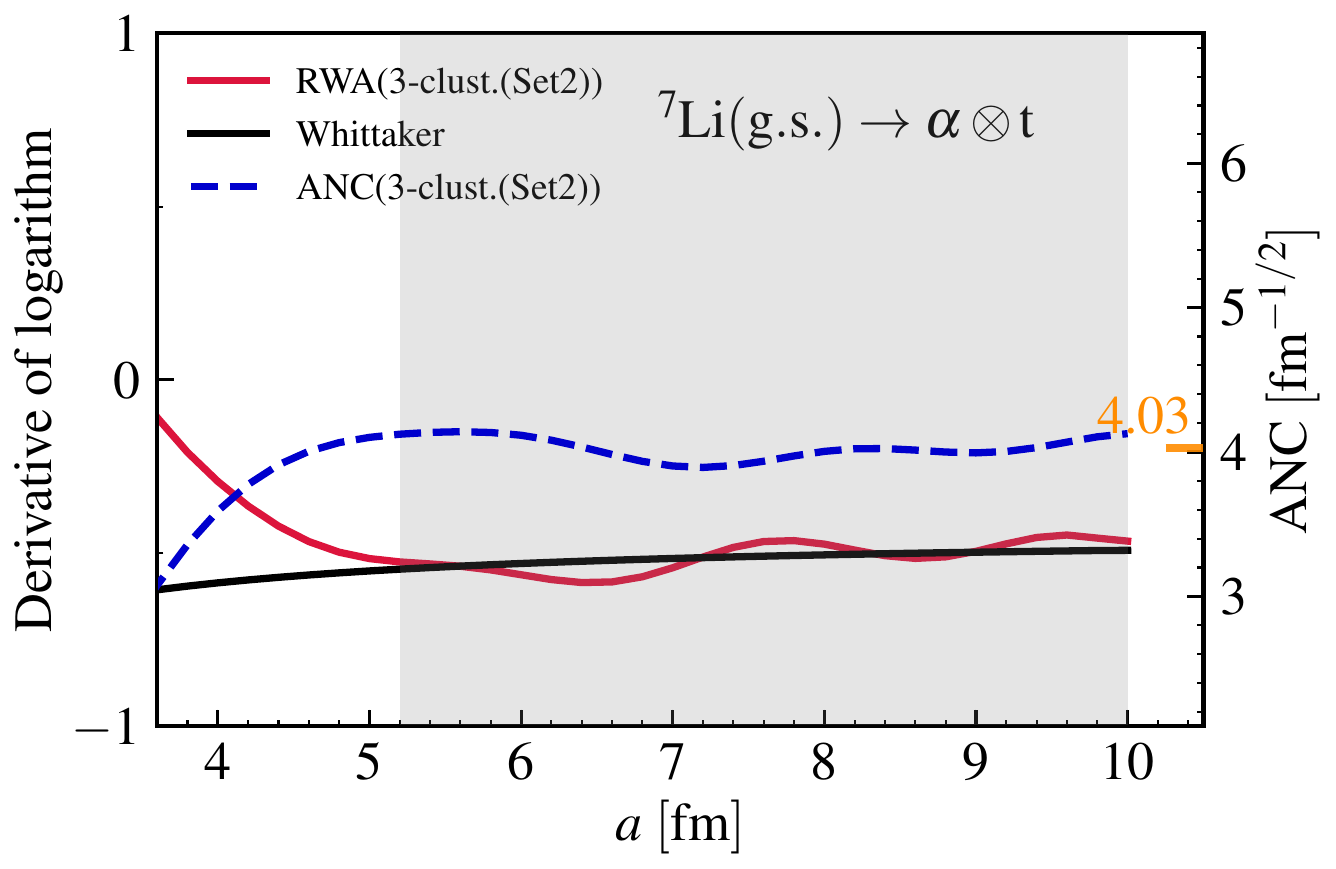}
  \end{subfigure}
  \hfill
  \begin{subfigure}[b]{0.48\textwidth}
    \includegraphics[width=\linewidth]{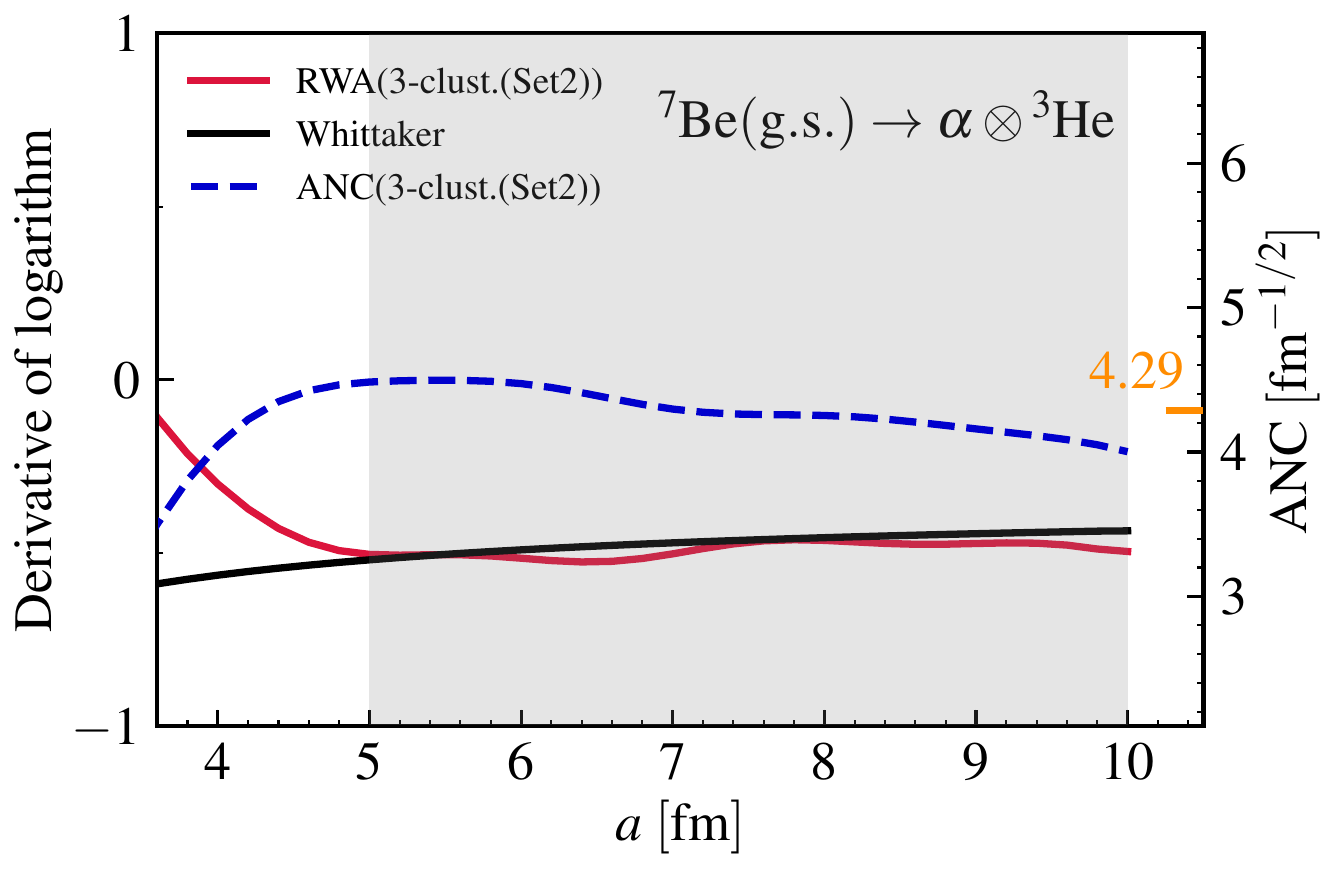}
  \end{subfigure}

  \caption{Comparison of the logarithmic derivatives of the RWA and Whittaker function (left axis) with the corresponding ANC value (right axis) under different sets. The ANC is calculated in the gray shaded region, with the orange short line indicating the mean value.}
  \label{fig:anc}
\end{figure*}
\begin{figure}
  \centering
  \begin{subfigure}[b]{0.48\textwidth}
    \includegraphics[width=\textwidth]{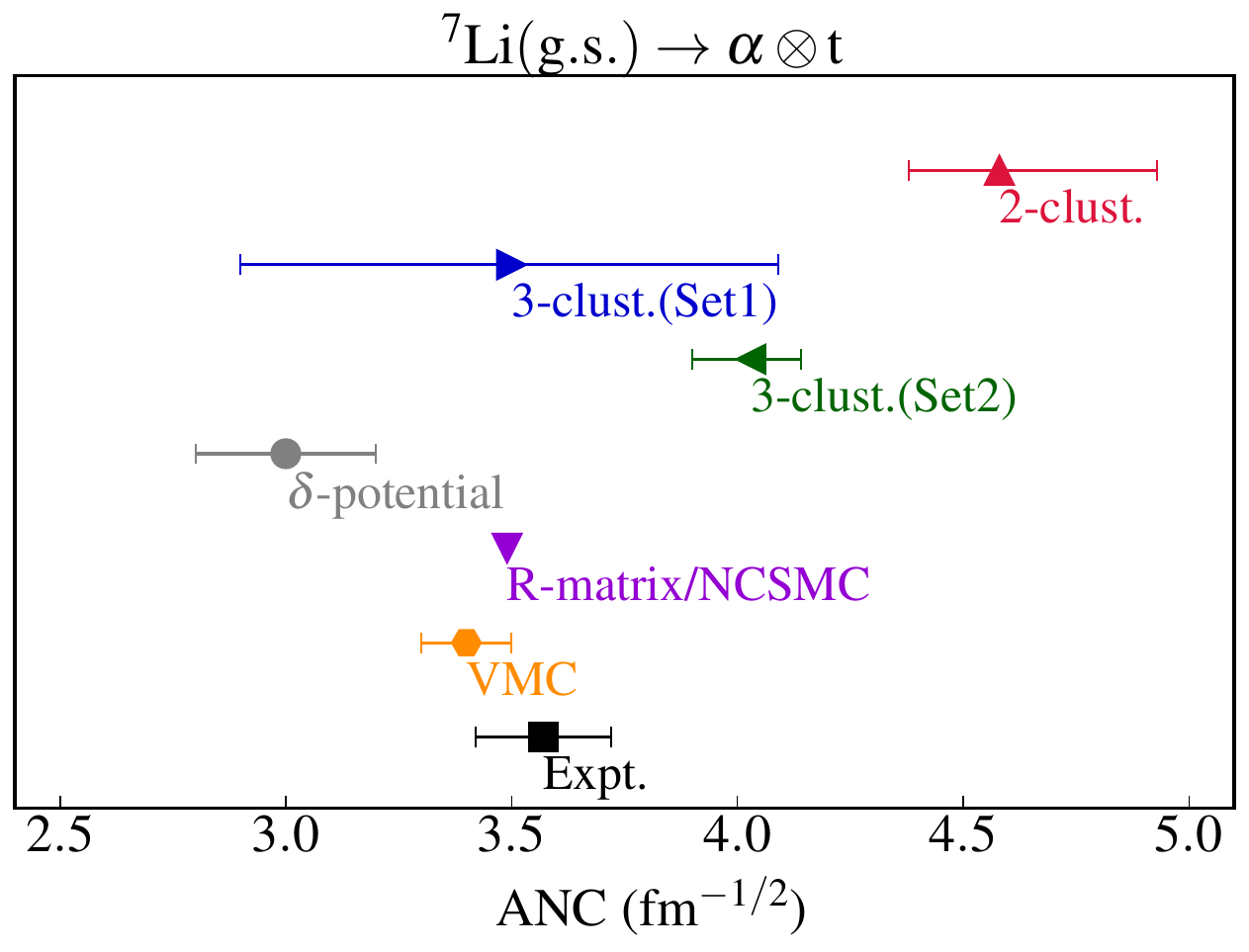}
  \end{subfigure}
  \\ 
  \vspace{0.5cm}
  \begin{subfigure}[b]{0.48\textwidth}
    \includegraphics[width=\textwidth]{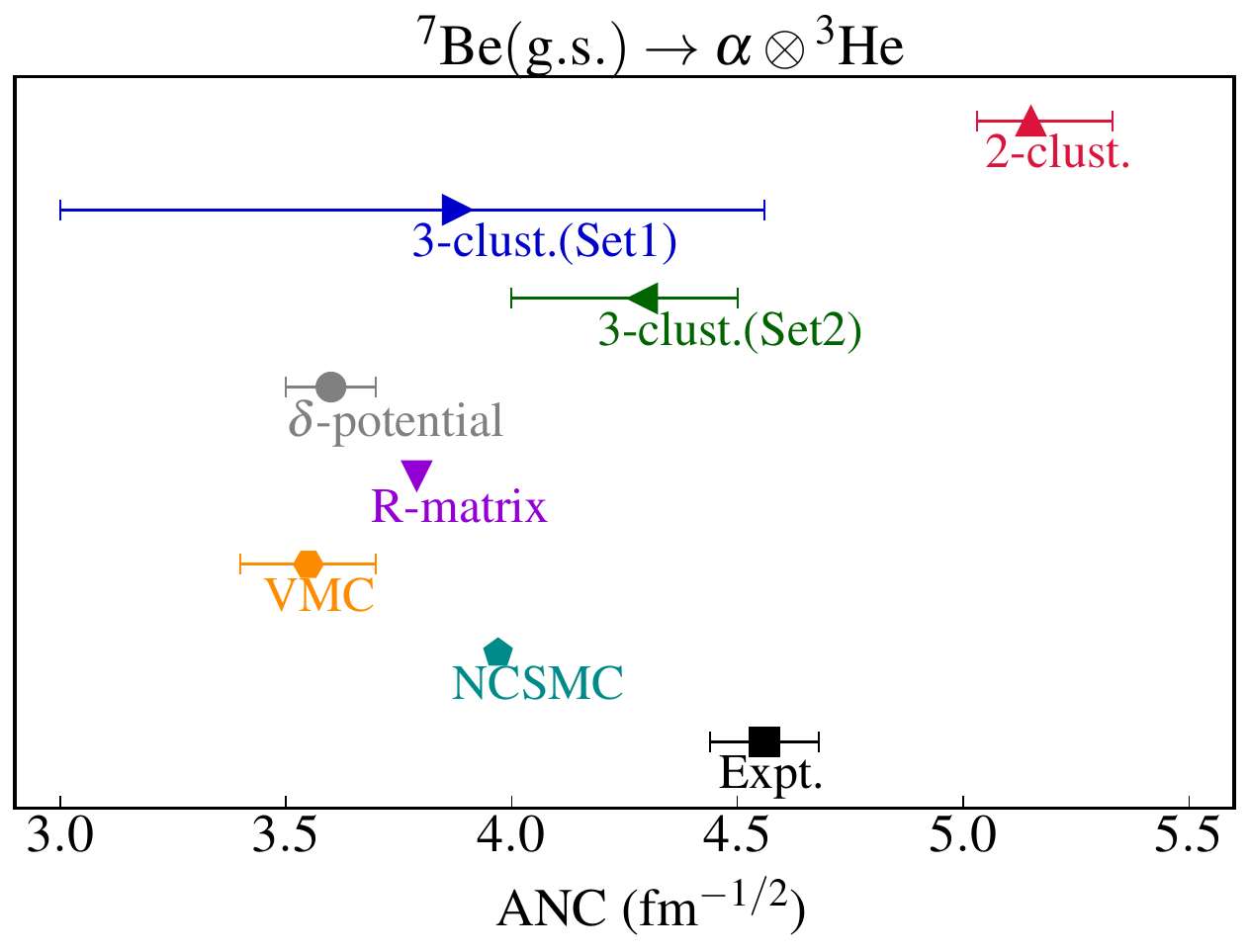}
  \end{subfigure}
  \caption{\label{fig:list}Summary of the most recent ANCs for $^7$Li(g.s.)$\rightarrow \alpha +t$ and $^7$Be(g.s.)$\rightarrow \alpha +{}^3\text{He}$ channels ($\mathrm{\delta}$-potential~\cite{PhysRevC.100.054307},R-matrix~\cite{DESCOUVEMONT2004203},NCSMC~\cite{PhysRevC.100.024304},VMC~\cite{PhysRevC.63.054002} and experimental data (Expt.)~\cite{KISS2020135606,IGAMOV2007247}) compared with present work(2-clust., 3-clust.(Set1), and 3-clust.(Set2)).}
\end{figure}

To illustrate the distinction in their asymptotic behavior, we compute their logarithmic derivatives following Eq.~(\ref{eq:dlog}). The results are shown in Fig.~\ref{fig:dlog}. The black line labeled ``Whittaker" represents the behavior of the Whittaker function, while the other results should be identical to it. It can be seen that, although the 2-clust. model exhibits shortcomings in energy and spectrum calculations, its asymptotic behavior performs remarkably well. We can imagine that if the limitation of insufficient model space in the 2-clust. model can be overcome, for instance, by calculating a nucleus with simpler structure such as $^8$Be, this model would serve as a proper simplified model for analyzing the ANC. Comparing the results of the two sets within the 3-clust. model, we observe that Set1 deviates from the Whittaker function compared to Set2. This is a dramatic result because we have demonstrated that the wave functions provided by these two sets are almost identical from common perspectives such as energy and overlap, but their asymptotic behavior exhibits significant differences. Next, we will reveal how this easily overlooked difference significantly impacts the extracted value of ANC.

\subsection{Asymptotic normalization coefficient}

In simple terms, ANC is the ratio of RWA to the Whittaker function in the tail region. It should be a constant theoretically, as long as the distance between residues is sufficiently large. Fig~\ref{fig:anc} simultaneously displays the variation of ANC with distance values and the logarithmic derivatives of the RWA and Whittaker functions.

Among these results, we can find that the ANC calculated from the 2-clust. and 3-clust. (Set2) models, which exhibit better asymptotic behavior, behave stably in the tail region. In this region, the ANC value can be approximately regarded as a constant. In contrast, due to the slight deviation in the asymptotic behavior of the 3-clust. (Set1) model from the Whitaker function, the calculated ANC from this model exhibits significant variations in the asymptotic region. This makes it difficult to precisely extract the ANC value from this result.

To extract the ANC value from the results in Fig~\ref{fig:anc}, we identify the first contact point $a_0$ as the starting point for our calculation. The ANC is then estimated as the mean of the values in the subsequent region $L~(L=10~\text{fm}-a_0)$(the gray shaded region in Fig~\ref{fig:anc}). The upper limit of $10$ fm is set because our diagonalization calculations become unstable beyond this distance. Besides, $10$ fm is much larger than the average size of a nucleus, a distance which is large enough for estimating the ANC. The uncertainty of the ANC value is determined by the maximum and minimum values within this region. We compare the estimated ANC results with those from several other theoretical or experimental approaches in Fig~\ref{fig:list}.

It can be seen that, first, although the results of the 2-clust. model are relatively stable, they clearly overestimate the ANC compared to other results. This indicates that the simple two-cluster model is not suitable for calculating the ANC of structurally complex nuclei such as $^7$Li and $^7$Be. When performing calculations using the 3-clust. model, we must also pay attention to the basis coverage in the model space at larger distances. Otherwise, we risk encountering extremely high uncertainties, as demonstrated by the results for 3-clust. (Set1) in the figure. After giving full consideration to the aforementioned issues concerning the model space, we obtained the results for the 3-clust. (Set2) model. It can be seen that these results agree well with experimental values, particularly for $^7$Be, where it aligns more closely with experimental expectations than other theoretical works. The numerical results are listed in Table~\ref{tab:summary}.
\begin{table}[htbp]
\caption{\label{table:summary}ANCs for $^7$Li(g.s.)$\rightarrow \alpha +t$ and $^7$Be(g.s.)$\rightarrow \alpha +{}^3\text{He}$ channels under different models.
}
\begin{ruledtabular}
\renewcommand{\arraystretch}{1.5}
\begin{tabular}{lcc}
\textrm{}&
\textrm{Structure}&
\multicolumn{1}{c}{ANC (fm${}^{-1/2}$)}\\
\colrule
$^7$Li & 2-clust.   & $4.58^{+0.35}_{-0.20}$ \\
       & 3-clust.(set1) & $3.50^{+0.59}_{-0.59}$ \\
       & 3-clust.(set2) & $4.03^{+0.13}_{-0.11}$ \\
$^7$Be & 2-clust.   & $5.15^{+0.18}_{-0.12}$\\
       & 3-clust.(set1) & $3.88^{+0.68}_{-0.88}$\\
       & 3-clust.(set2) & $4.29^{+0.21}_{-0.29}$\\
\end{tabular}
\end{ruledtabular}
\label{tab:summary}
\end{table}

\section{Summary}

In this work, we have performed a systematic study of the $^7$Li and $^7$Be nuclei and their ANCs within a microscopic cluster model framework. We compared results from a two-cluster model with those from a three-cluster model extended from it. Although the basis of the two-cluster model has a sufficiently broad spatial distribution, its model space is incomplete. The three-cluster model addresses this by augmenting the two-cluster basis with additional three-cluster configuration basis states. We implemented this extension using both a spatially compact and a diffuse set of three-cluster configurations to probe the sensitivity of the calculations to the description of the wave function's long-range behavior.

Our calculations show that both three-cluster models provide a more accurate description of the binding energies than the pure two-cluster model. However, a key distinction emerges in their asymptotic properties. The analysis reveals that while the model including only the compact three-cluster configurations is sufficient for describing the energy spectrum well, it fails to properly describe the asymptotic behavior of the RWA. The analysis of the RWA and its logarithmic derivative ultimately reveals that a proper description of the wave function's tail, achieved only with the addition of the diffuse basis, is crucial for obtaining reliable asymptotic behavior.

Consequently, the three-cluster model employing the diffuse three-cluster configurations yields ANCs for the ground states of $^7$Li and $^7$Be that are in good agreement with experimental data.

This study underscores two key conclusions for the theoretical determination of ANCs. First, for nuclei with complex internal structures, the use of multi-cluster or many-body models is essential for an accurate description. Second, and more critically, the configuration of the model space at large inter-cluster separations is paramount. We demonstrate that a component of the wave function constituting less than 0.3\% of the total norm can be decisive for the accurate and effective calculation of the ANC, even when its impact on the total energy is negligible.

\begin{acknowledgments}
This work was supported by National Natural Science Foundation of China [Grant Nos. 12305123, 12105141], by the Jiangsu Provincial Natural Science Foundation (Grants No. BK20210277), by the National Key R\&D Program of China (Contract No.2023YFA1606503), by the 2021 Jiangsu Shuangchuang (Mass Innovation and Entrepreneurship) Talent Program (Grants No.JSSCBS20210169). Numerical calculations were performed in the Cluster-Computing Center of School of Science (C3S2) at Huzhou University.
\end{acknowledgments}

\bibliography{apssamp}

\begin{thebibliography}{35}%
\makeatletter
\providecommand \@ifxundefined [1]{%
 \@ifx{#1\undefined}
}%
\providecommand \@ifnum [1]{%
 \ifnum #1\expandafter \@firstoftwo
 \else \expandafter \@secondoftwo
 \fi
}%
\providecommand \@ifx [1]{%
 \ifx #1\expandafter \@firstoftwo
 \else \expandafter \@secondoftwo
 \fi
}%
\providecommand \natexlab [1]{#1}%
\providecommand \enquote  [1]{``#1''}%
\providecommand \bibnamefont  [1]{#1}%
\providecommand \bibfnamefont [1]{#1}%
\providecommand \citenamefont [1]{#1}%
\providecommand \href@noop [0]{\@secondoftwo}%
\providecommand \href [0]{\begingroup \@sanitize@url \@href}%
\providecommand \@href[1]{\@@startlink{#1}\@@href}%
\providecommand \@@href[1]{\endgroup#1\@@endlink}%
\providecommand \@sanitize@url [0]{\catcode `\\12\catcode `\$12\catcode `\&12\catcode `\#12\catcode `\^12\catcode `\_12\catcode `\%12\relax}%
\providecommand \@@startlink[1]{}%
\providecommand \@@endlink[0]{}%
\providecommand \url  [0]{\begingroup\@sanitize@url \@url }%
\providecommand \@url [1]{\endgroup\@href {#1}{\urlprefix }}%
\providecommand \urlprefix  [0]{URL }%
\providecommand \Eprint [0]{\href }%
\providecommand \doibase [0]{https://doi.org/}%
\providecommand \selectlanguage [0]{\@gobble}%
\providecommand \bibinfo  [0]{\@secondoftwo}%
\providecommand \bibfield  [0]{\@secondoftwo}%
\providecommand \translation [1]{[#1]}%
\providecommand \BibitemOpen [0]{}%
\providecommand \bibitemStop [0]{}%
\providecommand \bibitemNoStop [0]{.\EOS\space}%
\providecommand \EOS [0]{\spacefactor3000\relax}%
\providecommand \BibitemShut  [1]{\csname bibitem#1\endcsname}%
\let\auto@bib@innerbib\@empty
\bibitem [{\citenamefont {Cyburt}\ \emph {et~al.}(2016)\citenamefont {Cyburt}, \citenamefont {Fields}, \citenamefont {Olive},\ and\ \citenamefont {Yeh}}]{RevModPhys.88.015004}%
  \BibitemOpen
  \bibfield  {author} {\bibinfo {author} {\bibfnamefont {R.~H.}\ \bibnamefont {Cyburt}}, \bibinfo {author} {\bibfnamefont {B.~D.}\ \bibnamefont {Fields}}, \bibinfo {author} {\bibfnamefont {K.~A.}\ \bibnamefont {Olive}},\ and\ \bibinfo {author} {\bibfnamefont {T.-H.}\ \bibnamefont {Yeh}},\ }\href {https://doi.org/10.1103/RevModPhys.88.015004} {\bibfield  {journal} {\bibinfo  {journal} {Rev. Mod. Phys.}\ }\textbf {\bibinfo {volume} {88}},\ \bibinfo {pages} {015004} (\bibinfo {year} {2016})}\BibitemShut {NoStop}%
\bibitem [{\citenamefont {Adelberger}\ \emph {et~al.}(1998)\citenamefont {Adelberger} \emph {et~al.}}]{RevModPhys.70.1265}%
  \BibitemOpen
  \bibfield  {author} {\bibinfo {author} {\bibfnamefont {E.~G.}\ \bibnamefont {Adelberger}} \emph {et~al.},\ }\href {https://doi.org/10.1103/RevModPhys.70.1265} {\bibfield  {journal} {\bibinfo  {journal} {Rev. Mod. Phys.}\ }\textbf {\bibinfo {volume} {70}},\ \bibinfo {pages} {1265} (\bibinfo {year} {1998})}\BibitemShut {NoStop}%
\bibitem [{\citenamefont {Nollett}\ and\ \citenamefont {Burles}(2000)}]{PhysRevD.61.123505}%
  \BibitemOpen
  \bibfield  {author} {\bibinfo {author} {\bibfnamefont {K.~M.}\ \bibnamefont {Nollett}}\ and\ \bibinfo {author} {\bibfnamefont {S.}~\bibnamefont {Burles}},\ }\href {https://doi.org/10.1103/PhysRevD.61.123505} {\bibfield  {journal} {\bibinfo  {journal} {Phys. Rev. D}\ }\textbf {\bibinfo {volume} {61}},\ \bibinfo {pages} {123505} (\bibinfo {year} {2000})}\BibitemShut {NoStop}%
\bibitem [{\citenamefont {Adelberger}\ \emph {et~al.}(2011)\citenamefont {Adelberger} \emph {et~al.}}]{RevModPhys.83.195}%
  \BibitemOpen
  \bibfield  {author} {\bibinfo {author} {\bibfnamefont {E.~G.}\ \bibnamefont {Adelberger}} \emph {et~al.},\ }\href {https://doi.org/10.1103/RevModPhys.83.195} {\bibfield  {journal} {\bibinfo  {journal} {Rev. Mod. Phys.}\ }\textbf {\bibinfo {volume} {83}},\ \bibinfo {pages} {195} (\bibinfo {year} {2011})}\BibitemShut {NoStop}%
\bibitem [{\citenamefont {Pizzone}\ \emph {et~al.}(2014)\citenamefont {Pizzone}, \citenamefont {Spartá}, \citenamefont {Bertulani}, \citenamefont {Spitaleri}, \citenamefont {La~Cognata}, \citenamefont {Lalmansingh}, \citenamefont {Lamia}, \citenamefont {Mukhamedzhanov},\ and\ \citenamefont {Tumino}}]{Pizzone_2014}%
  \BibitemOpen
  \bibfield  {author} {\bibinfo {author} {\bibfnamefont {R.~G.}\ \bibnamefont {Pizzone}}, \bibinfo {author} {\bibfnamefont {R.}~\bibnamefont {Spartá}}, \bibinfo {author} {\bibfnamefont {C.~A.}\ \bibnamefont {Bertulani}}, \bibinfo {author} {\bibfnamefont {C.}~\bibnamefont {Spitaleri}}, \bibinfo {author} {\bibfnamefont {M.}~\bibnamefont {La~Cognata}}, \bibinfo {author} {\bibfnamefont {J.}~\bibnamefont {Lalmansingh}}, \bibinfo {author} {\bibfnamefont {L.}~\bibnamefont {Lamia}}, \bibinfo {author} {\bibfnamefont {A.}~\bibnamefont {Mukhamedzhanov}},\ and\ \bibinfo {author} {\bibfnamefont {A.}~\bibnamefont {Tumino}},\ }\href {https://doi.org/10.1088/0004-637X/786/2/112} {\bibfield  {journal} {\bibinfo  {journal} {Astrophys. J.}\ }\textbf {\bibinfo {volume} {786}},\ \bibinfo {pages} {112} (\bibinfo {year} {2014})}\BibitemShut {NoStop}%
\bibitem [{\citenamefont {Christy}\ and\ \citenamefont {Duck}(1961)}]{CHRISTY196189}%
  \BibitemOpen
  \bibfield  {author} {\bibinfo {author} {\bibfnamefont {R.~F.}\ \bibnamefont {Christy}}\ and\ \bibinfo {author} {\bibfnamefont {I.}~\bibnamefont {Duck}},\ }\href {https://doi.org/https://doi.org/10.1016/0029-5582(61)91019-7} {\bibfield  {journal} {\bibinfo  {journal} {Nucl. Phys.}\ }\textbf {\bibinfo {volume} {24}},\ \bibinfo {pages} {89} (\bibinfo {year} {1961})}\BibitemShut {NoStop}%
\bibitem [{\citenamefont {Xu}\ \emph {et~al.}(1994)\citenamefont {Xu}, \citenamefont {Gagliardi}, \citenamefont {Tribble}, \citenamefont {Mukhamedzhanov},\ and\ \citenamefont {Timofeyuk}}]{PhysRevLett.73.2027}%
  \BibitemOpen
  \bibfield  {author} {\bibinfo {author} {\bibfnamefont {H.~M.}\ \bibnamefont {Xu}}, \bibinfo {author} {\bibfnamefont {C.~A.}\ \bibnamefont {Gagliardi}}, \bibinfo {author} {\bibfnamefont {R.~E.}\ \bibnamefont {Tribble}}, \bibinfo {author} {\bibfnamefont {A.~M.}\ \bibnamefont {Mukhamedzhanov}},\ and\ \bibinfo {author} {\bibfnamefont {N.~K.}\ \bibnamefont {Timofeyuk}},\ }\href {https://doi.org/10.1103/PhysRevLett.73.2027} {\bibfield  {journal} {\bibinfo  {journal} {Phys. Rev. Lett.}\ }\textbf {\bibinfo {volume} {73}},\ \bibinfo {pages} {2027} (\bibinfo {year} {1994})}\BibitemShut {NoStop}%
\bibitem [{\citenamefont {Mukhamedzhanov}\ \emph {et~al.}(2003)\citenamefont {Mukhamedzhanov}, \citenamefont {B{\'e}m}, \citenamefont {Brown}, \citenamefont {Burjan}, \citenamefont {Gagliardi}, \citenamefont {Kroha}, \citenamefont {Nov{\'a}k}, \citenamefont {Nunes}, \citenamefont {Pisko{\v{r}}}, \citenamefont {Pirlepesov}, \citenamefont {{\v{S}}ime{\v{c}}kov{\'a}}, \citenamefont {Tribble},\ and\ \citenamefont {Vincour}}]{PhysRevC.67.065804}%
  \BibitemOpen
  \bibfield  {author} {\bibinfo {author} {\bibfnamefont {A.~M.}\ \bibnamefont {Mukhamedzhanov}}, \bibinfo {author} {\bibfnamefont {P.}~\bibnamefont {B{\'e}m}}, \bibinfo {author} {\bibfnamefont {B.~A.}\ \bibnamefont {Brown}}, \bibinfo {author} {\bibfnamefont {V.}~\bibnamefont {Burjan}}, \bibinfo {author} {\bibfnamefont {C.~A.}\ \bibnamefont {Gagliardi}}, \bibinfo {author} {\bibfnamefont {V.}~\bibnamefont {Kroha}}, \bibinfo {author} {\bibfnamefont {J.}~\bibnamefont {Nov{\'a}k}}, \bibinfo {author} {\bibfnamefont {F.~M.}\ \bibnamefont {Nunes}}, \bibinfo {author} {\bibfnamefont {{\v{S}}.}~\bibnamefont {Pisko{\v{r}}}}, \bibinfo {author} {\bibfnamefont {F.}~\bibnamefont {Pirlepesov}}, \bibinfo {author} {\bibfnamefont {E.}~\bibnamefont {{\v{S}}ime{\v{c}}kov{\'a}}}, \bibinfo {author} {\bibfnamefont {R.~E.}\ \bibnamefont {Tribble}},\ and\ \bibinfo {author} {\bibfnamefont {J.}~\bibnamefont {Vincour}},\ }\href {https://doi.org/10.1103/PhysRevC.67.065804} {\bibfield  {journal} {\bibinfo  {journal} {Phys. Rev. C}\ }\textbf
  {\bibinfo {volume} {67}},\ \bibinfo {pages} {065804} (\bibinfo {year} {2003})}\BibitemShut {NoStop}%
\bibitem [{\citenamefont {Mohr}\ \emph {et~al.}(1993)\citenamefont {Mohr}, \citenamefont {Abele}, \citenamefont {Zwiebel}, \citenamefont {Staudt}, \citenamefont {Krauss}, \citenamefont {Oberhummer}, \citenamefont {Denker}, \citenamefont {Hammer},\ and\ \citenamefont {Wolf}}]{PhysRevC.48.1420}%
  \BibitemOpen
  \bibfield  {author} {\bibinfo {author} {\bibfnamefont {P.}~\bibnamefont {Mohr}}, \bibinfo {author} {\bibfnamefont {H.}~\bibnamefont {Abele}}, \bibinfo {author} {\bibfnamefont {R.}~\bibnamefont {Zwiebel}}, \bibinfo {author} {\bibfnamefont {G.}~\bibnamefont {Staudt}}, \bibinfo {author} {\bibfnamefont {H.}~\bibnamefont {Krauss}}, \bibinfo {author} {\bibfnamefont {H.}~\bibnamefont {Oberhummer}}, \bibinfo {author} {\bibfnamefont {A.}~\bibnamefont {Denker}}, \bibinfo {author} {\bibfnamefont {J.~W.}\ \bibnamefont {Hammer}},\ and\ \bibinfo {author} {\bibfnamefont {G.}~\bibnamefont {Wolf}},\ }\href {https://doi.org/10.1103/PhysRevC.48.1420} {\bibfield  {journal} {\bibinfo  {journal} {Phys. Rev. C}\ }\textbf {\bibinfo {volume} {48}},\ \bibinfo {pages} {1420} (\bibinfo {year} {1993})}\BibitemShut {NoStop}%
\bibitem [{\citenamefont {Mohr}(2009)}]{PhysRevC.79.065804}%
  \BibitemOpen
  \bibfield  {author} {\bibinfo {author} {\bibfnamefont {P.}~\bibnamefont {Mohr}},\ }\href {https://doi.org/10.1103/PhysRevC.79.065804} {\bibfield  {journal} {\bibinfo  {journal} {Phys. Rev. C}\ }\textbf {\bibinfo {volume} {79}},\ \bibinfo {pages} {065804} (\bibinfo {year} {2009})}\BibitemShut {NoStop}%
\bibitem [{\citenamefont {{Dubovichenko}}\ and\ \citenamefont {{Dzhazairov-Kakhramanov}}(1995)}]{1995PAN....58..579D}%
  \BibitemOpen
  \bibfield  {author} {\bibinfo {author} {\bibfnamefont {S.~B.}\ \bibnamefont {{Dubovichenko}}}\ and\ \bibinfo {author} {\bibfnamefont {A.~V.}\ \bibnamefont {{Dzhazairov-Kakhramanov}}},\ }\href {https://doi.org/10.48550/arXiv.nucl-th/9802080} {\bibfield  {journal} {\bibinfo  {journal} {Phys. At. Nucl.}\ }\textbf {\bibinfo {volume} {58}},\ \bibinfo {pages} {579} (\bibinfo {year} {1995})},\ \Eprint {https://arxiv.org/abs/nucl-th/9802080} {arXiv:nucl-th/9802080 [nucl-th]} \BibitemShut {NoStop}%
\bibitem [{\citenamefont {Liu}\ \emph {et~al.}(1981)\citenamefont {Liu}, \citenamefont {Kanada},\ and\ \citenamefont {Tang}}]{PhysRevC.23.645}%
  \BibitemOpen
  \bibfield  {author} {\bibinfo {author} {\bibfnamefont {Q.~K.~K.}\ \bibnamefont {Liu}}, \bibinfo {author} {\bibfnamefont {H.}~\bibnamefont {Kanada}},\ and\ \bibinfo {author} {\bibfnamefont {Y.~C.}\ \bibnamefont {Tang}},\ }\href {https://doi.org/10.1103/PhysRevC.23.645} {\bibfield  {journal} {\bibinfo  {journal} {Phys. Rev. C}\ }\textbf {\bibinfo {volume} {23}},\ \bibinfo {pages} {645} (\bibinfo {year} {1981})}\BibitemShut {NoStop}%
\bibitem [{\citenamefont {Langanke}(1986)}]{LANGANKE1986351}%
  \BibitemOpen
  \bibfield  {author} {\bibinfo {author} {\bibfnamefont {K.}~\bibnamefont {Langanke}},\ }\href {https://doi.org/10.1016/0375-9474(86)90383-0} {\bibfield  {journal} {\bibinfo  {journal} {Nucl. Phys. A}\ }\textbf {\bibinfo {volume} {457}},\ \bibinfo {pages} {351} (\bibinfo {year} {1986})}\BibitemShut {NoStop}%
\bibitem [{\citenamefont {Kajino}(1986)}]{KAJINO1986559}%
  \BibitemOpen
  \bibfield  {author} {\bibinfo {author} {\bibfnamefont {T.}~\bibnamefont {Kajino}},\ }\href {https://doi.org/10.1016/0375-9474(86)90428-8} {\bibfield  {journal} {\bibinfo  {journal} {Nucl. Phys. A}\ }\textbf {\bibinfo {volume} {460}},\ \bibinfo {pages} {559} (\bibinfo {year} {1986})}\BibitemShut {NoStop}%
\bibitem [{\citenamefont {Mertelmeier}\ and\ \citenamefont {Hofmann}(1986)}]{MERTELMEIER1986387}%
  \BibitemOpen
  \bibfield  {author} {\bibinfo {author} {\bibfnamefont {T.}~\bibnamefont {Mertelmeier}}\ and\ \bibinfo {author} {\bibfnamefont {H.~M.}\ \bibnamefont {Hofmann}},\ }\href {https://doi.org/10.1016/0375-9474(86)90141-7} {\bibfield  {journal} {\bibinfo  {journal} {Nucl. Phys. A}\ }\textbf {\bibinfo {volume} {459}},\ \bibinfo {pages} {387} (\bibinfo {year} {1986})}\BibitemShut {NoStop}%
\bibitem [{\citenamefont {Cs{\'o}t{\'o}}\ and\ \citenamefont {Langanke}(2000)}]{csoto2000study}%
  \BibitemOpen
  \bibfield  {author} {\bibinfo {author} {\bibfnamefont {A.}~\bibnamefont {Cs{\'o}t{\'o}}}\ and\ \bibinfo {author} {\bibfnamefont {K.}~\bibnamefont {Langanke}},\ }\href {https://doi.org/https://doi.org/10.1007/s006010070012} {\bibfield  {journal} {\bibinfo  {journal} {Few-Body Syst.}\ }\textbf {\bibinfo {volume} {29}},\ \bibinfo {pages} {121} (\bibinfo {year} {2000})}\BibitemShut {NoStop}%
\bibitem [{\citenamefont {Timofeyuk}\ \emph {et~al.}(2008)\citenamefont {Timofeyuk}, \citenamefont {Thompson},\ and\ \citenamefont {Tostevin}}]{NKTimofeyuk_2008}%
  \BibitemOpen
  \bibfield  {author} {\bibinfo {author} {\bibfnamefont {N.~K.}\ \bibnamefont {Timofeyuk}}, \bibinfo {author} {\bibfnamefont {I.~J.}\ \bibnamefont {Thompson}},\ and\ \bibinfo {author} {\bibfnamefont {J.~A.}\ \bibnamefont {Tostevin}},\ }\href {https://doi.org/10.1088/1742-6596/111/1/012034} {\bibfield  {journal} {\bibinfo  {journal} {J. Phys.: Conf. Ser.}\ }\textbf {\bibinfo {volume} {111}},\ \bibinfo {pages} {012034} (\bibinfo {year} {2008})}\BibitemShut {NoStop}%
\bibitem [{\citenamefont {Timofeyuk}(2014)}]{Timofeyuk_2014}%
  \BibitemOpen
  \bibfield  {author} {\bibinfo {author} {\bibfnamefont {N.~K.}\ \bibnamefont {Timofeyuk}},\ }\href {https://doi.org/10.1088/0954-3899/41/9/094008} {\bibfield  {journal} {\bibinfo  {journal} {J. Phys. G}\ }\textbf {\bibinfo {volume} {41}},\ \bibinfo {pages} {094008} (\bibinfo {year} {2014})}\BibitemShut {NoStop}%
\bibitem [{\citenamefont {Myo}\ \emph {et~al.}(2024)\citenamefont {Myo}, \citenamefont {Lyu}, \citenamefont {Zhao}, \citenamefont {Isaka}, \citenamefont {Wan}, \citenamefont {Takemoto}, \citenamefont {Horiuchi},\ and\ \citenamefont {Doté}}]{10.1093/ptep/ptae187}%
  \BibitemOpen
  \bibfield  {author} {\bibinfo {author} {\bibfnamefont {T.}~\bibnamefont {Myo}}, \bibinfo {author} {\bibfnamefont {M.}~\bibnamefont {Lyu}}, \bibinfo {author} {\bibfnamefont {Q.}~\bibnamefont {Zhao}}, \bibinfo {author} {\bibfnamefont {M.}~\bibnamefont {Isaka}}, \bibinfo {author} {\bibfnamefont {N.}~\bibnamefont {Wan}}, \bibinfo {author} {\bibfnamefont {H.}~\bibnamefont {Takemoto}}, \bibinfo {author} {\bibfnamefont {H.}~\bibnamefont {Horiuchi}},\ and\ \bibinfo {author} {\bibfnamefont {A.}~\bibnamefont {Doté}},\ }\href {https://doi.org/10.1093/ptep/ptae187} {\bibfield  {journal} {\bibinfo  {journal} {Prog. Theor. Exp. Phys.}\ }\textbf {\bibinfo {volume} {2025}},\ \bibinfo {pages} {013D01} (\bibinfo {year} {2024})}\BibitemShut {NoStop}%
\bibitem [{\citenamefont {Volkov}(1965)}]{VOLKOV196533}%
  \BibitemOpen
  \bibfield  {author} {\bibinfo {author} {\bibfnamefont {A.~B.}\ \bibnamefont {Volkov}},\ }\href {https://doi.org/10.1016/0029-5582(65)90244-0} {\bibfield  {journal} {\bibinfo  {journal} {Nucl. Phys.}\ }\textbf {\bibinfo {volume} {74}},\ \bibinfo {pages} {33} (\bibinfo {year} {1965})}\BibitemShut {NoStop}%
\bibitem [{\citenamefont {Tamagaki}(1968)}]{10.1143/PTP.39.91}%
  \BibitemOpen
  \bibfield  {author} {\bibinfo {author} {\bibfnamefont {R.}~\bibnamefont {Tamagaki}},\ }\href {https://doi.org/10.1143/PTP.39.91} {\bibfield  {journal} {\bibinfo  {journal} {Prog. Theor. Phys.}\ }\textbf {\bibinfo {volume} {39}},\ \bibinfo {pages} {91} (\bibinfo {year} {1968})}\BibitemShut {NoStop}%
\bibitem [{\citenamefont {Yamaguchi}\ \emph {et~al.}(1979)\citenamefont {Yamaguchi}, \citenamefont {Kasahara}, \citenamefont {Nagata},\ and\ \citenamefont {Akaishi}}]{10.1143/PTP.62.1018}%
  \BibitemOpen
  \bibfield  {author} {\bibinfo {author} {\bibfnamefont {N.}~\bibnamefont {Yamaguchi}}, \bibinfo {author} {\bibfnamefont {T.}~\bibnamefont {Kasahara}}, \bibinfo {author} {\bibfnamefont {S.}~\bibnamefont {Nagata}},\ and\ \bibinfo {author} {\bibfnamefont {Y.}~\bibnamefont {Akaishi}},\ }\href {https://doi.org/10.1143/PTP.62.1018} {\bibfield  {journal} {\bibinfo  {journal} {Prog. Theor. Phys.}\ }\textbf {\bibinfo {volume} {62}},\ \bibinfo {pages} {1018} (\bibinfo {year} {1979})}\BibitemShut {NoStop}%
\bibitem [{\citenamefont {Cheng}\ \emph {et~al.}(2025)\citenamefont {Cheng}, \citenamefont {Lyu}, \citenamefont {Myo}, \citenamefont {Horiuchi}, \citenamefont {Toki}, \citenamefont {Ren}, \citenamefont {Isaka}, \citenamefont {Mao}, \citenamefont {Takemoto}, \citenamefont {Wan}, \citenamefont {You},\ and\ \citenamefont {Zhao}}]{CHENG2025139397}%
  \BibitemOpen
  \bibfield  {author} {\bibinfo {author} {\bibfnamefont {Z.}~\bibnamefont {Cheng}}, \bibinfo {author} {\bibfnamefont {M.}~\bibnamefont {Lyu}}, \bibinfo {author} {\bibfnamefont {T.}~\bibnamefont {Myo}}, \bibinfo {author} {\bibfnamefont {H.}~\bibnamefont {Horiuchi}}, \bibinfo {author} {\bibfnamefont {H.}~\bibnamefont {Toki}}, \bibinfo {author} {\bibfnamefont {Z.}~\bibnamefont {Ren}}, \bibinfo {author} {\bibfnamefont {M.}~\bibnamefont {Isaka}}, \bibinfo {author} {\bibfnamefont {M.}~\bibnamefont {Mao}}, \bibinfo {author} {\bibfnamefont {H.}~\bibnamefont {Takemoto}}, \bibinfo {author} {\bibfnamefont {N.}~\bibnamefont {Wan}}, \bibinfo {author} {\bibfnamefont {W.}~\bibnamefont {You}},\ and\ \bibinfo {author} {\bibfnamefont {Q.}~\bibnamefont {Zhao}},\ }\href {https://doi.org/https://doi.org/10.1016/j.physletb.2025.139397} {\bibfield  {journal} {\bibinfo  {journal} {Phys. Lett. B}\ }\textbf {\bibinfo {volume} {864}},\ \bibinfo {pages} {139397} (\bibinfo {year} {2025})}\BibitemShut {NoStop}%
\bibitem [{\citenamefont {Tian}\ \emph {et~al.}(2024)\citenamefont {Tian}, \citenamefont {Cheng}, \citenamefont {Yu}, \citenamefont {Lyu}, \citenamefont {Myo}, \citenamefont {Isaka}, \citenamefont {Toki}, \citenamefont {Horiuchi}, \citenamefont {Doté}, \citenamefont {Takemoto}, \citenamefont {Wan},\ and\ \citenamefont {Zhao}}]{TIAN2024138816}%
  \BibitemOpen
  \bibfield  {author} {\bibinfo {author} {\bibfnamefont {J.}~\bibnamefont {Tian}}, \bibinfo {author} {\bibfnamefont {Z.}~\bibnamefont {Cheng}}, \bibinfo {author} {\bibfnamefont {C.}~\bibnamefont {Yu}}, \bibinfo {author} {\bibfnamefont {M.}~\bibnamefont {Lyu}}, \bibinfo {author} {\bibfnamefont {T.}~\bibnamefont {Myo}}, \bibinfo {author} {\bibfnamefont {M.}~\bibnamefont {Isaka}}, \bibinfo {author} {\bibfnamefont {H.}~\bibnamefont {Toki}}, \bibinfo {author} {\bibfnamefont {H.}~\bibnamefont {Horiuchi}}, \bibinfo {author} {\bibfnamefont {A.}~\bibnamefont {Doté}}, \bibinfo {author} {\bibfnamefont {H.}~\bibnamefont {Takemoto}}, \bibinfo {author} {\bibfnamefont {N.}~\bibnamefont {Wan}},\ and\ \bibinfo {author} {\bibfnamefont {Q.}~\bibnamefont {Zhao}},\ }\href {https://doi.org/https://doi.org/10.1016/j.physletb.2024.03745} {\bibfield  {journal} {\bibinfo  {journal} {Phys. Lett. B}\ }\textbf {\bibinfo {volume} {855}},\ \bibinfo {pages} {138816} (\bibinfo {year} {2024})}\BibitemShut {NoStop}%
\bibitem [{\citenamefont {Chiba}\ and\ \citenamefont {Kimura}(2017)}]{10.1093/ptep/ptx063}%
  \BibitemOpen
  \bibfield  {author} {\bibinfo {author} {\bibfnamefont {Y.}~\bibnamefont {Chiba}}\ and\ \bibinfo {author} {\bibfnamefont {M.}~\bibnamefont {Kimura}},\ }\href {https://doi.org/10.1093/ptep/ptx063} {\bibfield  {journal} {\bibinfo  {journal} {Prog. Theor. Exp. Phys.}\ }\textbf {\bibinfo {volume} {2017}},\ \bibinfo {pages} {053D01} (\bibinfo {year} {2017})}\BibitemShut {NoStop}%
\bibitem [{\citenamefont {Zhao}\ \emph {et~al.}(2022)\citenamefont {Zhao}, \citenamefont {Kimura}, \citenamefont {Zhou},\ and\ \citenamefont {Shin}}]{PhysRevC.106.054313}%
  \BibitemOpen
  \bibfield  {author} {\bibinfo {author} {\bibfnamefont {Q.}~\bibnamefont {Zhao}}, \bibinfo {author} {\bibfnamefont {M.}~\bibnamefont {Kimura}}, \bibinfo {author} {\bibfnamefont {B.}~\bibnamefont {Zhou}},\ and\ \bibinfo {author} {\bibfnamefont {S.-h.}\ \bibnamefont {Shin}},\ }\href {https://doi.org/10.1103/PhysRevC.106.054313} {\bibfield  {journal} {\bibinfo  {journal} {Phys. Rev. C}\ }\textbf {\bibinfo {volume} {106}},\ \bibinfo {pages} {054313} (\bibinfo {year} {2022})}\BibitemShut {NoStop}%
\bibitem [{\citenamefont {Purcell}\ and\ \citenamefont {Sheu}(2015)}]{PURCELL20151}%
  \BibitemOpen
  \bibfield  {author} {\bibinfo {author} {\bibfnamefont {J.~E.}\ \bibnamefont {Purcell}}\ and\ \bibinfo {author} {\bibfnamefont {C.~G.}\ \bibnamefont {Sheu}},\ }\href {https://doi.org/https://doi.org/10.1016/j.nds.2015.11.001} {\bibfield  {journal} {\bibinfo  {journal} {Nucl. Data Sheets}\ }\textbf {\bibinfo {volume} {130}},\ \bibinfo {pages} {1} (\bibinfo {year} {2015})}\BibitemShut {NoStop}%
\bibitem [{\citenamefont {Tilley}\ \emph {et~al.}(1992)\citenamefont {Tilley}, \citenamefont {Weller},\ and\ \citenamefont {Hale}}]{TILLEY19921}%
  \BibitemOpen
  \bibfield  {author} {\bibinfo {author} {\bibfnamefont {D.~R.}\ \bibnamefont {Tilley}}, \bibinfo {author} {\bibfnamefont {H.~R.}\ \bibnamefont {Weller}},\ and\ \bibinfo {author} {\bibfnamefont {G.~M.}\ \bibnamefont {Hale}},\ }\href {https://doi.org/https://doi.org/10.1016/0375-9474(92)90635-W} {\bibfield  {journal} {\bibinfo  {journal} {Nucl. Phys. A}\ }\textbf {\bibinfo {volume} {541}},\ \bibinfo {pages} {1} (\bibinfo {year} {1992})}\BibitemShut {NoStop}%
\bibitem [{\citenamefont {Tilley}\ \emph {et~al.}(2002)\citenamefont {Tilley}, \citenamefont {Cheves}, \citenamefont {Godwin}, \citenamefont {Hale}, \citenamefont {Hofmann}, \citenamefont {Kelley}, \citenamefont {Sheu},\ and\ \citenamefont {Weller}}]{TILLEY20023}%
  \BibitemOpen
  \bibfield  {author} {\bibinfo {author} {\bibfnamefont {D.~R.}\ \bibnamefont {Tilley}}, \bibinfo {author} {\bibfnamefont {C.~M.}\ \bibnamefont {Cheves}}, \bibinfo {author} {\bibfnamefont {J.~L.}\ \bibnamefont {Godwin}}, \bibinfo {author} {\bibfnamefont {G.~M.}\ \bibnamefont {Hale}}, \bibinfo {author} {\bibfnamefont {H.~M.}\ \bibnamefont {Hofmann}}, \bibinfo {author} {\bibfnamefont {J.~H.}\ \bibnamefont {Kelley}}, \bibinfo {author} {\bibfnamefont {C.~G.}\ \bibnamefont {Sheu}},\ and\ \bibinfo {author} {\bibfnamefont {H.~R.}\ \bibnamefont {Weller}},\ }\href {https://doi.org/https://doi.org/10.1016/S0375-9474(02)00597-3} {\bibfield  {journal} {\bibinfo  {journal} {Nucl. Phys. A}\ }\textbf {\bibinfo {volume} {708}},\ \bibinfo {pages} {3} (\bibinfo {year} {2002})}\BibitemShut {NoStop}%
\bibitem [{\citenamefont {Luna}\ and\ \citenamefont {Papenbrock}(2019)}]{PhysRevC.100.054307}%
  \BibitemOpen
  \bibfield  {author} {\bibinfo {author} {\bibfnamefont {B.~K.}\ \bibnamefont {Luna}}\ and\ \bibinfo {author} {\bibfnamefont {T.}~\bibnamefont {Papenbrock}},\ }\href {https://doi.org/10.1103/PhysRevC.100.054307} {\bibfield  {journal} {\bibinfo  {journal} {Phys. Rev. C}\ }\textbf {\bibinfo {volume} {100}},\ \bibinfo {pages} {054307} (\bibinfo {year} {2019})}\BibitemShut {NoStop}%
\bibitem [{\citenamefont {Descouvemont}\ \emph {et~al.}(2004)\citenamefont {Descouvemont}, \citenamefont {Adahchour}, \citenamefont {Angulo}, \citenamefont {Coc},\ and\ \citenamefont {Vangioni-Flam}}]{DESCOUVEMONT2004203}%
  \BibitemOpen
  \bibfield  {author} {\bibinfo {author} {\bibfnamefont {P.}~\bibnamefont {Descouvemont}}, \bibinfo {author} {\bibfnamefont {A.}~\bibnamefont {Adahchour}}, \bibinfo {author} {\bibfnamefont {C.}~\bibnamefont {Angulo}}, \bibinfo {author} {\bibfnamefont {A.}~\bibnamefont {Coc}},\ and\ \bibinfo {author} {\bibfnamefont {E.}~\bibnamefont {Vangioni-Flam}},\ }\href {https://doi.org/https://doi.org/10.1016/j.adt.2004.08.001} {\bibfield  {journal} {\bibinfo  {journal} {At. Data Nucl. Data Tables}\ }\textbf {\bibinfo {volume} {88}},\ \bibinfo {pages} {203} (\bibinfo {year} {2004})}\BibitemShut {NoStop}%
\bibitem [{\citenamefont {Vorabbi}\ \emph {et~al.}(2019)\citenamefont {Vorabbi}, \citenamefont {Navr\'atil}, \citenamefont {Quaglioni},\ and\ \citenamefont {Hupin}}]{PhysRevC.100.024304}%
  \BibitemOpen
  \bibfield  {author} {\bibinfo {author} {\bibfnamefont {M.}~\bibnamefont {Vorabbi}}, \bibinfo {author} {\bibfnamefont {P.}~\bibnamefont {Navr\'atil}}, \bibinfo {author} {\bibfnamefont {S.}~\bibnamefont {Quaglioni}},\ and\ \bibinfo {author} {\bibfnamefont {G.}~\bibnamefont {Hupin}},\ }\href {https://doi.org/10.1103/PhysRevC.100.024304} {\bibfield  {journal} {\bibinfo  {journal} {Phys. Rev. C}\ }\textbf {\bibinfo {volume} {100}},\ \bibinfo {pages} {024304} (\bibinfo {year} {2019})}\BibitemShut {NoStop}%
\bibitem [{\citenamefont {Nollett}(2001)}]{PhysRevC.63.054002}%
  \BibitemOpen
  \bibfield  {author} {\bibinfo {author} {\bibfnamefont {K.~M.}\ \bibnamefont {Nollett}},\ }\href {https://doi.org/10.1103/PhysRevC.63.054002} {\bibfield  {journal} {\bibinfo  {journal} {Phys. Rev. C}\ }\textbf {\bibinfo {volume} {63}},\ \bibinfo {pages} {054002} (\bibinfo {year} {2001})}\BibitemShut {NoStop}%
\bibitem [{\citenamefont {Kiss}\ \emph {et~al.}(2020)\citenamefont {Kiss} \emph {et~al.}}]{KISS2020135606}%
  \BibitemOpen
  \bibfield  {author} {\bibinfo {author} {\bibfnamefont {G.}~\bibnamefont {Kiss}} \emph {et~al.},\ }\href {https://doi.org/https://doi.org/10.1016/j.physletb.2020.135606} {\bibfield  {journal} {\bibinfo  {journal} {Physics Letters B}\ }\textbf {\bibinfo {volume} {807}},\ \bibinfo {pages} {135606} (\bibinfo {year} {2020})}\BibitemShut {NoStop}%
\bibitem [{\citenamefont {Igamov}\ and\ \citenamefont {Yarmukhamedov}(2007)}]{IGAMOV2007247}%
  \BibitemOpen
  \bibfield  {author} {\bibinfo {author} {\bibfnamefont {S.}~\bibnamefont {Igamov}}\ and\ \bibinfo {author} {\bibfnamefont {R.}~\bibnamefont {Yarmukhamedov}},\ }\href {https://doi.org/https://doi.org/10.1016/j.nuclphysa.2006.10.041} {\bibfield  {journal} {\bibinfo  {journal} {Nuclear Physics A}\ }\textbf {\bibinfo {volume} {781}},\ \bibinfo {pages} {247} (\bibinfo {year} {2007})}\BibitemShut {NoStop}%
\end{thebibliography}%

\end{document}